\documentclass[twocolumn]{autart}    
\usepackage{amsmath} 
\usepackage{hyperref}
\usepackage{xcolor}

\newcommand{\bb}{\color{black}}
\newcommand{\tb}{\color{black}}

\usepackage{multirow} 
\usepackage{amsfonts} 
\usepackage{amssymb} 
\usepackage{amsmath}
\usepackage{booktabs}
\usepackage{graphicx}          
\newtheorem{problem}{Problem}[section]
\newtheorem{theorem}{Theorem}[section]
\newtheorem{lemma}{Lemma}[section]
\newtheorem{myremark}{Remark}[section]
\newtheorem{assumption}{Assumption}
\newenvironment{remark}{\begin{myremark}\normalfont}{\end{myremark}}
\newenvironment{proof}{\vspace{.1cm}\noindent{\sc Proof.}\hspace{0.10cm}\,\,}{$\hfill\Box$\vspace{.1cm}} 
\newcommand{\rline}{{\mathbb R}}

\begin{document}

\begin{frontmatter}

\title{Distributed end-effector formation control for mixed fully- and under-actuated manipulators with flexible joints\thanksref{footnoteinfo}} 

\thanks[footnoteinfo]{This paper was not presented at any IFAC 
meeting. Corresponding author Z.~Peng.}

\author[Nanjing,Groningen]{Zhiyu Peng}\ead{230208667@seu.edu.cn},    
\author[Groningen]{Bayu Jayawardhana}\ead{b.jayawardhana@rug.nl},   
\author[Nanjing]{Xin Xin}\ead{101012606@seu.edu.cn}  

\address[Nanjing]{Southeast University, Nanjing}
\address[Groningen]{University of Groningen, Groningen}


\begin{abstract}                          
The presence of faulty or underactuated manipulators can disrupt the end-effector formation keeping of a team of manipulators. Based on two-link planar manipulators, we investigate this end-effector formation keeping problem for mixed fully- and under-actuated manipulators with flexible joints. In this case, the underactuated manipulators can comprise of active-passive (AP) manipulators, passive-active (PA) manipulators, or a combination thereof. We propose distributed control laws for the different types of manipulators to achieve and maintain the desired formation shape of the end-effectors. It is achieved by assigning virtual springs to the end-effectors for the fully-actuated ones and to the virtual end-effectors for the under-actuated ones. We further study the set of all desired and reachable shapes for the networked manipulators’ end-effectors. Finally, we validate our analysis via numerical simulations. 
\end{abstract}

\end{frontmatter}

\section{Introduction}
\label{sec:introduction}
There has been a growing interest in cooperative control of manipulators, enabling a group of manipulators to jointly execute complex tasks \cite{chen2019distributed,liu2023adaptive,wu2021distributed,wu2022distributed}. Recently, Wu et al. \cite{wu2021distributed,wu2022distributed} explore the design of distributed end-effector formation control laws for a team of fully-actuated manipulators (which possess an equal number of inputs and degrees of freedom). In \cite{wu2021distributed,wu2022distributed}, two distributed control design methods are proposed: the distance-based and the displacement-based method; for a detailed discussion on these methods, we refer interested readers to  \cite{oh2015survey}. The basic idea of the control strategies in \cite{wu2021distributed,wu2022distributed} is to apply virtual springs to connect the edges of the formation graph of the manipulators' end-effectors.

However, when some manipulators in the group are underactuated (i.e., they have fewer inputs than degrees of freedom), the control strategies in \cite{wu2021distributed,wu2022distributed} are no longer applicable. 
The presence of underactuated manipulators may be caused by faulty actuation in some of the joints.  
While the control of a single underactuated manipulator has been well-studied for the past decades (see, e.g. \cite{fantoni2000energy,zhang2013motion,chen2021controllability,FL01,XL14}), the distributed cooperative control of multi underactuated manipulators remains underdeveloped. In particular, it remains an open problem in the design of distributed end-effector formation controllers for a combination of fully- and under-actuated manipulators. 

Using planar two-link manipulators, this paper investigates the end-effector formation keeping control where some manipulators are underactuated with a single active actuator. Depending on which joint is actuated, we consider either the active-passive (AP) manipulators, where the active actuator is in the first joint, the passive-active (PA) manipulators with the active actuator in the second joint, or both. To simplify the problem formulation, the manipulators are assumed to operate in the same gravity-free plane. 

The complexity of this control problem arises primarily from the second-order nonholonomic constraint of the underactuated manipulator \cite{arai1998nonholonomic,oriolo1991control}, which implies that the angle of the single-actuated joint cannot fully determine the end-effector position. In our previous work \cite{peng2023distributed}, we developed distributed end-effector formation controllers for networked two-link manipulators operating in a gravity-free plane, consisting of fully-actuated ones and underactuated PA manipulators. The distributed controller proposed for the PA manipulator in \cite{peng2023distributed} relies on the integrability of its nonholonomic constraint \cite{lai2015stabilization,oriolo1991control}. However, the method has the following limitations. Firstly, it only applies to the PA manipulator with an initial joint velocity of zero. {\bb Secondly, the controller design requires precise information about mechanical characteristics such as the center of mass (CoM) and the moment of inertia. 
} Thirdly, the presence of small disturbances, such as joint friction, can already render the nonholonomic constraint non-integrable and the controller ineffective.

In this paper, to tackle the end-effector formation-keeping problem, we introduce a torsional spring at each manipulator's joint (see also Fig. \ref{fig1}). The adaptation of the mechanical structure, in conjunction with a novel control strategy, allows us to overcome the aforementioned limitations effectively. The novel distributed control strategy uses 
a {\bb defined} virtual end-effector position of {\bb the 
}local manipulator, which is determined solely by its active joint angle(s). For fully-actuated manipulators, this virtual position coincides with the actual end-effector position. For underactuated ones, this virtual position is defined by setting the active joint angle to its current value and the passive joint angle to zero. {\bb As shown later, the introduced torsional springs establish a relationship between the virtual and actual positions of end-effectors.} We refer interested readers to the control of underactuated manipulators with flexible joints (i.e., joints with torsional springs) to \cite{XL13ASME,peng2023energy,zhang2017nonlinear,wu2023control}.  

The main contributions of this paper are as follows. Firstly, we briefly review on 
relevant properties of 
underactuated manipulators with flexible joints (consisting of the AP and PA manipulators). One particular property of interest is that if the active joint angle is constant under a constant control torque, then the underactuated manipulator is at the equilibrium point with its passive joint angle being zero.
This property implies that when the manipulator is in a steady state, its virtual and actual end-effector positions are the same. {\bb For the AP manipulator,
} while this property is validated in \cite[Chapter 8]{FL01}, we provide a simpler proof. 
 {\bb For the PA manipulator, 
}we prove that this property always holds except for a particular set of mechanical parameters.

Secondly, we extend the control strategies in \cite{wu2021distributed,wu2022distributed} to the aforementioned group of {\bb manipulators with flexible joints.}
In contrast to the approach pursued in \cite{wu2021distributed,wu2022distributed}, we employ virtual springs to couple the defined virtual end-effector positions for the underactuated agents, instead of the actual ones. Using these defined virtual positions as intermediates, we design distributed controllers and provide a stability analysis for closed-loop systems. The stability analysis shows that the networked actual end-effectors converge to the desired formation shape, assuming the manipulators avoid a set of singular points.

Thirdly, we discuss the set of desired and reachable shapes $\mathcal{S}_W$ for the networked manipulators’ end-effectors. Our study reveals an inverse relationship between the number of underactuated manipulators in the group and {\bb the cardinality of $\mathcal{S}_W$}. Moreover, a comparative study between distance-based and displacement-based methods in terms of {\bb the cardinality of $\mathcal{S}_W$} reveals the advantages of the former method. For the group with three or fewer underactuated manipulators, the distance-based method is effective, while the displacement-based method is effective for two or fewer underactuated agents. We validate our proposed method and analysis numerically. 

The remainder of this paper is organized as follows. In Section \ref{sec:preliminaries}, we give a preliminary on the dynamics and kinematics of the manipulators, the basic distributed formation control theory, and the formulation of the main problem. In Section \ref{main result}, we detail our main results. Finally, we present simulations and conclusions in Sections \ref{sec:sim} and \ref{sec:con}, respectively.

\section{Preliminaries and Problem Formulation}
\label{sec:preliminaries}

\emph{Notation:} For a set of column vectors $x_i$ and a set of matrices $A_i$, $i=1,...,N$, let ${{\mathop{\rm col}\nolimits} _{i \in \{1,...,N\} }}\left( { \ldots ,{x_i}, \ldots } \right): = {\mathop{\rm col}\nolimits} \left( {{x_1}, \ldots ,{x_N}} \right) \vspace{0.1cm}= {\left[ {x_1^{\rm{T}}, \ldots ,x_N^{\rm{T}}} \right]^{\rm{T}}}$ be the
stacked column vector and  $\operatorname{block\;diag} _{i \in \{ 1,...,N\} }\left(\ldots, A_i, \ldots\right) $ be a block diagonal matrix of $A_i$. Let $I_N \in \mathbb{R}^{N \times N}$ denote the identity matrix, $\mathbf{1}_N \in \mathbb{R}^N$ the vector composed of all ones, and \(\otimes\) the Kronecker product symbol. 

\begin{figure}
	\centering
	\includegraphics[scale=0.45]{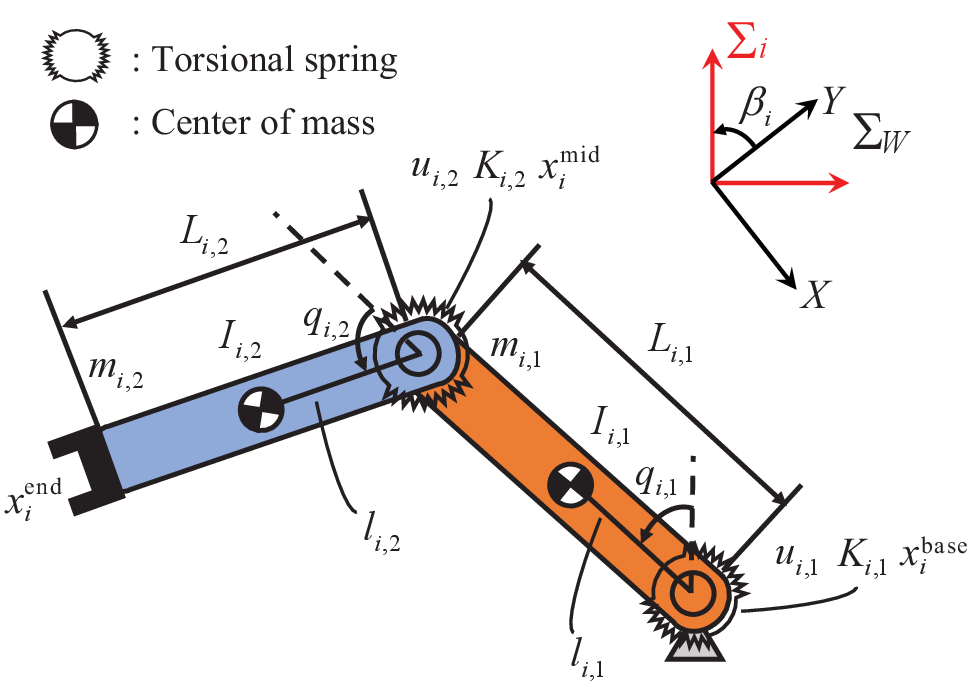}
	\caption{The planar two-link manipulator with flexible joints. If both joints are active, it is a fully-actuated manipulator. In this paper, we call it the AP manipulator if its first joint (the base joint) is active and its second joint (the joint connecting two links) is passive; or, we call it the PA manipulator if it has a passive first joint and an active second joint.} 
 \label{fig1}
\end{figure}

This paper focuses on the networked $N$ two-link manipulators with flexible joints (Fig. \ref{fig1}) operating in the same gravity-free $X-Y$ plane. 
For the $j$-th ($j = 1, 2$) link of manipulator $i$, the symbols $m_{i,j}$, $I_{i,j}$, $L_{i,j}$, and $l_{i,j}$ represent its mass, the moment of inertia concerning
its CoM, its length, and the distance from the $j$-th joint to its CoM, respectively. For each joint $j$, a torsional spring is attached, with $K_{i,j}$ representing the stiffness. Denote $\beta_i$ as the relative orientation angle between the world (global) coordinate frame $\Sigma_W$ and the local coordinate frame $\Sigma_i$. The vectors $x_i^{{\rm{base}}}: = \left[ {\begin{array}{*{20}{c}}
{x_{i,X}^{{\rm{base}}}}, &{x_{i,Y}^{{\rm{base}}}}
\end{array}} \right]^{\rm T}$, $x_i^{{\rm{mid}}}: = \left[ {\begin{array}{*{20}{c}}
{x_{i,X}^{{\rm{mid}}}}, &{x_{i,Y}^{{\rm{mid}}}}
\end{array}} \right]^{\rm T}$, and $x_i^{{\rm{end}}}: = \left[ {\begin{array}{*{20}{c}}
{x_{i,X}^{{\rm{end}}}}, &{x_{i,Y}^{{\rm{end}}}}
\end{array}} \right]^{\rm T}$ denote the positions of the fixed base, the mid-joint, and the end-effector within $\Sigma_W$, respectively.

\subsection{Manipulator Dynamics and Kinematics}
In order to differentiate the types of manipulators, the following three sets are defined 
\[
\begin{aligned}
& \mathcal{M}_{\mathrm{fa}}:=\{i: \text { manipulator } i \text { is fully-actuated}\}, \\
& \mathcal{M}_{\mathrm{ap}}:=\{i: \text { manipulator } i \text { is the AP manipulator}\}, \\
& \mathcal{M}_{\mathrm{pa}}:=\{i: \text { manipulator } i \text { is the PA manipulator}\},
\end{aligned}
\]
where the subscripts `fa', `ap', and `pa' refer to the fully-actuated, AP, and PA manipulators respectively. Note that the sets $\mathcal{M}_{\mathrm{fa}},\; \mathcal{M}_{\mathrm{ap}}$ and $\mathcal{M}_{\mathrm{pa}}$ have no intersection with each other and $\mathcal{M}: = {\mathcal{M}_{{\rm{fa}}}} \cup {\mathcal{M}_{{\rm{ap}}}} \cup {\mathcal{M}_{{\rm{pa}}}} = \left\{ {1,...,N} \right\}$. 
Consider that the numbers of fully-actuated, AP and PA manipulators are given by $|{\mathcal{M}_{{\rm{fa}}}}|=N_1$, $|{\mathcal{M}_{{\rm{ap}}}}|=N_2 $ and $|{\mathcal{M}_{{\rm{pa}}}}|=N_3 $, respectively, such that $N_1+N_2+N_3=N$. Without loss of generality, in the subsequent sections, we assume that 
${\mathcal{M}_{{\rm{fa}}}} = \{ i\in \mathbb{Z}_{+}:1 \le i \le {N_1}\}$, 
${\mathcal{M}_{{\rm{ap}}}} = \{ i\in \mathbb{Z}_{+}:N_1+1 \le i \le {N_1+N_2}\}$, and 
${\mathcal{M}_{{\rm{pa}}}} = \{ i \in {\mathbb{Z}_+ }:{N_1} + {N_2} + 1 \le i \le N\} $. Here, \(\mathbb{Z}_+\) represents the set of all positive integers.

Using the Euler-Lagrange equation \cite{murray2017mathematical,spong2006robot}, we can describe the motion of the mentioned manipulators by 
\begin{equation}
	M_i\left(q_i\right) \ddot{q}_i+C_i\left(q_i, \dot{q}_i\right) \dot{q}_i + K_iq_i= u_i, \; i \in \mathcal{M},\label{full}
\end{equation}
where $q_i=[q_{i,1}, q_{i,2}]^\mathrm{T}$ is the joint angle and $u_i=\left[u_{i, 1}, u_{i, 2}\right]^{\mathrm{T}}$ denotes the input. For all $i \in \mathcal{M}_\text{ap}$, we have $u_{i,2} = 0$, indicating that its second joint is passive; similarly, $u_{i,1} = 0$ for all $i \in \mathcal{M}_\text{pa}$, signifying the first joint being passive. The term $K_i = \left(\begin{smallmatrix}K_{i,1} & 0 \\ 0 & K_{i,2}\end{smallmatrix}\right)$ is the stiffness matrix of the 
torsional spring at the joints. The mass matrix $M_i(q_i)$, which is positive definite, and the Coriolis and centrifugal term $C_i({q_i},{{\dot q}_i})$ are respectively given by 
\[\begin{aligned}
M_i\left(q_i\right) & =\left[\begin{array}{lc}
M_{i,11}\left(q_i\right) & M_{i,12}\left(q_i\right) \\
M_{i,21}\left(q_i\right) & M_{i,22}\left(q_i\right)
\end{array}\right] \\
 = & \left[\begin{array}{cc}
\alpha_{i,1}+\alpha_{i,2}+2 \alpha_{i,3} \cos q_{i,2} & \alpha_{i,2}+\alpha_{i,3} \cos q_{i,2} \\
\alpha_{i,2}+\alpha_{i,3} \cos q_{i,2} & \alpha_{i,2}
\end{array}\right],
\end{aligned}\]
\[C_i({q_i},{{\dot q}_i}) = {\alpha _{i,3}}\left[ {\begin{array}{*{20}{c}}
		{ - {{\dot q}_{i,2}}\;\;}&{ - {{\dot q}_{i,1}} - {{\dot q}_{i,2}}}\\
		{{{\dot q}_{i,1}}}&0
\end{array}} \right]\sin {q_{i,2}},\]
where $\alpha_{i,1}=m_{i,1} l_{i,1}^2+m_{i,2} L_{i,1}^2+I_{i,1}$, $\alpha_{i,2}=m_{i,2} l_{i,2}^2+I_{i,2}$, and $\alpha_{i,3}=m_{i,2} L_{i,1} l_{i,2}$ are the mechanical parameters. Following standard properties of Euler-Lagrange systems \cite{spong2006robot,Ortega-book}, the matrix $\dot M_i(q_i)-2C_i(q_i, \dot q_i)$ is skew-symmetric. 

Since $M_i(q_i)$ is positive definite, we can rewrite (\ref{full}) compactly into 
\begin{equation}
	{\ddot q = {M^{ - 1}}(q)\Big(u - C(q,\dot q)\dot q-Kq\Big)}, \label{dd}
\end{equation}
where the stacked column vectors $q,u \in \mathbb{R}^{2N}$ are given by $q : = {{\mathop{\rm col}\nolimits} _{i \in \mathcal{M}}}\left( { \ldots ,{q_i}, \ldots } \right)$ and $u : = {\mathop{\rm col}\nolimits} _{i \in \mathcal{M}}\left(\ldots ,{u_i}, \ldots  \right)$. The matrices $ M(q)$, $C(q,\dot q)$, $K \in \mathbb{R}^{2N\times2N}$ are block diagonal matrices of $ M_i(q_i), C_i(q_i,\dot q_i)$ and $K_i$ for every $i \in \mathcal{M}$, respectively.

For each manipulator $i$, the end-effector position $x_i^\text{end} \in \mathbb{R}^2$ can be expressed by 
\begin{equation}
\begin{aligned}
& x_i^{\text {end }}=h_i\left(q_{i, 1}, q_{i, 2}\right)+x_i^{\text {base }}, \\
& h_i\left(q_{i, 1}, q_{i, 2}\right): \mathbb{R}^2 \rightarrow \mathbb{R}^2 \\
& =\left[\begin{array}{c}
-L_{i, 1} \sin \left(q_{i, 1}+\beta_i\right)-L_{i, 2} \sin \left(q_{i, 1}+q_{i, 2}+\beta_i\right) \\
L_{i, 1} \cos \left(q_{i, 1}+\beta_i\right)+L_{i, 2} \cos \left(q_{i, 1}+q_{i, 2}+\beta_i\right)
\end{array}\right] .
\end{aligned} \label{task-space} 
\end{equation}
Accordingly, the time-derivative of \eqref{task-space} satisfies
\begin{equation}
\begin{array}{l}
{{\dot x}_i}^\text{end} = {J_i}\left( {{q_i}} \right){{\dot q}_i}, \\
{J_i}\left( {{q_i}} \right): = \displaystyle \frac{{\partial {h_i}}}{{\partial {q_i}}}\; = \left[ {\begin{array}{*{20}{l}}
{{J_{i,11}}\left( {{q_i}} \right)}&{{J_{i,12}}\left( {{q_i}} \right)}\\
{{J_{i,21}}\left( {{q_i}} \right)}&{{J_{i,22}}\left( {{q_i}} \right)}
\end{array}} \right],
\end{array}
\label{jaco}
\end{equation}
where $J_i(q_i) \in \mathbb{R}^{2 \times 2}$ is the standard Jacobian matrix of the forward kinematics  \cite{murray2017mathematical,spong2006robot} whose specific expression is omitted for brevity. 

Before proceeding to the subsequent section, we introduce additional notations for stacked vectors and diagonal matrices associated with the manipulator sets $\mathcal M_{\text{fa}}, \mathcal M_{\text{ap}}$ and $\mathcal M_{\text{pa}}$. 
To distinguish types of joints, the superscript 
`a' and `u' are used to refer to the actuated/active and unactuated/passive joints, respectively. 
The stack vectors of all active and of all passive (unactuated) joint angles are given by 
\begin{equation}
\begin{aligned}
& q^{\mathrm{a}}:=\operatorname{col}\left(q_{\mathrm{fa}}, q_{\mathrm{ap}}^{\mathrm{a}}, q_{\mathrm{pa}}^{\mathrm{a}}\right) \in \mathbb{R}^{N+N_1}, \\
& q^{\mathrm{u}}:=\operatorname{col}\left(q_{\mathrm{ap}}^{\mathrm{u}}, q_{\mathrm{pa}}^{\mathrm{u}}\right) \in \mathbb{R}^{N_2+N_3},
\end{aligned}  \label{qa}
\end{equation}
where the vector $q_{\mathrm{fa}} \in \mathbb{R}^{2N_1}$ denotes the stacked vector of all joint angles of fully-actuated manipulators, defined as $q_{\mathrm{fa}}:=\operatorname{col}_{i \in \mathcal{M}_{\mathrm{fa}}}\left(\ldots, q_i, \ldots\right)$; the vectors $q_{\mathrm{ap}}^{\mathrm{a}} \in \mathbb{R}^{N_2}$, $q_{\mathrm{pa}}^{\mathrm{a}} \in \mathbb{R}^{N_3}$ correspond to the stacked vectors of active joint angles of AP and PA manipulators respectively, formulated as 
$q_{\mathrm{ap}}^{\mathrm{a}}:=\operatorname{col}_{i \in \mathcal{M}_{\mathrm{ap}}}\left(\ldots, q_{i, 1}, \ldots\right)$, $q_{\mathrm{pa}}^{\mathrm{a}}:=\operatorname{col}_{i \in \mathcal{M}_{\mathrm{pa}}}\left(\ldots, q_{i, 2}, \ldots\right)$; similarly, the vectors $q_{\mathrm{ap}}^{\mathrm{u}} \in \mathbb{R}^{N_2}$, $q_{\mathrm{pa}}^{\mathrm{u}} \in \mathbb{R}^{N_3}$ are defined as $q_{\mathrm{ap}}^{\mathrm{u}}:=\operatorname{col}_{i \in \mathcal{M}_{\mathrm{ap}}}\left(\ldots, q_{i, 2}, \ldots\right)$, $q_{\mathrm{pa}}^{\mathrm{u}}:=\operatorname{col}_{i \in \mathcal{M}_{\mathrm{pa}}}\left(\ldots, q_{i, 1}, \ldots\right)$.

Correspondingly, the combined stiffness matrices of the torsional springs, attached at the active and passive joints, are respectively given by 
\begin{equation}
\begin{aligned}
& K^\mathrm{a}:=\operatorname{block\;diag}\left(K_{\mathrm{fa}}, K_{\mathrm{ap}}^{\mathrm{a}}, K_{\mathrm{pa}}^{\mathrm{a}}\right)\in \mathbb{R}^{({N+N_1}) \times ({N+N_1})}, \\
& K^{\mathrm{u}}:=\operatorname{block\;diag}\left(K_{\mathrm{ap}}^{\mathrm{u}}, K_{\mathrm{pa}}^{\mathrm{u}}\right)\in \mathbb{R}^{(N_2+N_3) \times (N_2+N_3)}, 
\end{aligned}
\label{defi:ka}
\end{equation}
where $K_{\mathrm{fa}}:=\operatorname{block\;diag}_{i \in \mathcal{M}_{\mathrm{fa}}}\left(\ldots, K_i, \ldots\right) \in \mathbb{R}^{2 N_1 \times 2 N_1}$, $ K_{\mathrm{ap}}^{\mathrm{a}}:=\operatorname{block\;diag}_{i \in \mathcal{M}_{\mathrm{ap}}}\left(\ldots, K_{i, 1}, \ldots\right) \in \mathbb{R}^{N_2 \times N_2}$, $K_{\mathrm{ap}}^{\mathrm{u}}:=\operatorname{block\;diag}_{i \in \mathcal{M}_{\mathrm{ap}}}\left(\ldots, K_{i, 2}, \ldots\right) \in \mathbb{R}^{N_2 \times N_2}$, $K_{\mathrm{pa}}^{\mathrm{a}}:=\operatorname{block\;diag}_{i \in \mathcal{M}_{\mathrm{pa}}}\left(\ldots, K_{i, 2}, \ldots\right) \in \mathbb{R}^{N_3 \times N_3}$, $K_{\mathrm{pa}}^{\mathrm{u}}:=\operatorname{block\;diag}_{i \in \mathcal{M}_{\mathrm{pa}}}\left(\ldots, K_{i, 1}, \ldots\right) \in \mathbb{R}^{N_3 \times N_3}$.

\subsection{Formation Graph and Problem Description}
\label{sec:graph}
In this subsection, we will review the use of an undirected graph $\mathcal{G}:=(\mathcal{V}, \mathcal{E})$ for defining a desired formation shape of the manipulators' end-effectors and for designing the corresponding distributed controllers. For the graph $\mathcal G$, denote $\mathcal{V} = \mathcal{M} = \{1,...,N\}$ as the vertex set and $\mathcal{E} \subset \mathcal{V} \times \mathcal{V}$ as the ordered edge set with $\mathcal{E}_k$ denoting the $k$-th edge. Let ${\mathcal{N}_i}: = \{ j \in \mathcal{V}:(i,j) \in \mathcal{E}\}$ be the set of the neighbors of the agent $i$. As usual, we denote the cardinality of $\mathcal V$ and $\mathcal E$ by 
$|\mathcal{V}| = N$ and $|\mathcal{E}|$. 
Associated to $\mathcal G$, define the elements of the incidence matrix $B \in \mathbb{R}^{N \times|\mathcal{E}|}$ by 
\begin{equation}
b_{i k}=\left\{\begin{array}{ll}
+1, & i=\mathcal{E}_k^{\text {tail }} \\
-1, & i=\mathcal{E}_k^{\text {head }} \\
0, & \text {otherwise}
\end{array} \; i = 1, \ldots,N, \; k=1, \ldots,|\mathcal{E}|,\right.
\end{equation}
where $\mathcal{E}_k^{\text {tail }}$ and $\mathcal{E}_k^{\text {head}}$ denote the tail and head nodes of $\mathcal{E}_k$ respectively, i.e., $(\mathcal{E}_k^{\text {tail }}, \mathcal{E}_k^{\text {head}})=\mathcal{E}_k$. Denote the stacked end-effector position vector  $x^\text{end} \in \mathbb{R}^{2N}:=\operatorname{col}_{i \in \mathcal{M}}\left(\ldots, x_i^\text{end}, \ldots \right)$. 
For a given reference end-effector position $x^* \in \mathbb{R}^{2N}$ that forms a desired formation shape, the set of all admissible positions with the desired shape is given by 
\begin{equation}
\resizebox{\linewidth}{!}{$\mathcal{S}:=\left\{x^\text{end}: x^\text{end}=\left(I_N \otimes R\right) x^* +\mathbf{1}_N \otimes c, R \in \mathbf{S O}(2), c \in \mathbb{R}^2\right\},$} \label{S}
\end{equation}
i.e., all admissible positions are obtained by applying translation and rotation to $x^*$. Let $\mathcal{S}_W \subset \mathcal{S}$ be the set of desired and reachable shapes for the networked manipulators' end-effectors, and we are now prepared to formulate the formation keeping problem discussed in this paper.

\begin{problem}
\label{problem}
{\bf (The Distributed End-Effector Formation Control Problem of Mixed Planar Fully- and Under-actuated Manipulator with Flexible Joints)} 
Consider the aforementioned group of planar two-link manipulators with flexible joints, including both fully- and under-actuated agents. 
For suitable fixed $x_{i}^{\rm base}, \beta_i,i=1,...,N$, and a given desired formation shape defined by the framework $(\mathcal G,x^*)$, design the distributed controller of the form
\begin{equation}
u_i = \sigma(\{x_j^{\rm end}, x_{j}^{\rm mid}\}_{j\in \mathcal N_i},q_i,\dot q_i), \;\; i=1,...,N,  \label{form}
\end{equation}
such that $x^{\rm end}(t) \to \mathcal{S}_W$ and $\dot q(t) \to {\bf 0}$ as $t \to \infty$. 
\end{problem} 

Observe that for any underactuated manipulator $i \in \mathcal{M}_{\mathrm{ap}}$, $u_{i,2}$ is always zero, while for any $i \in \mathcal{M}_{\mathrm{pa}}$, $u_{i,1}$ is always zero. 
Following the distributed controller form in \eqref{form} and as will be shown in our main results later, 
if the manipulator ${j\in \mathcal N_i}$ is underactuated, then the information of $x_{j}^\text{mid}$ is also needed for the control design. This information can be captured via a motion tracking camera that can track not only the movement of the end-effector $x_{j}^\text{end}$, but also that of the mid-joint $x_{j}^\text{mid}$. This extra information is a design trade-off due to the underactuation. It is in contrast to the existing results for the distributed end-effector formation control of fully-actuated manipulators, where the distributed controllers are in the form of   
\begin{equation}
u_i = \sigma(\{x_j^\text{end}\}_{j\in \mathcal N_i},q_i,\dot q_i), \;\; i=1,...,N
\end{equation}
(c.f. recent results in  \cite{wu2021distributed,wu2022distributed}).

{\bb To solve Problem {\ref{problem}}, this paper uses two prominent methods in formation control, namely the distance-based method \cite{oh2014distance} and the displacement-based method \cite{de2020maneuvering}. According to \cite{oh2015survey}, the desired formation for the manipulators' end-effectors can be defined by the constraint 
\begin{equation}
    f_{\mathcal{G}}(x^\text{end})=f_{\mathcal{G}}(x^*),
    \label{constraint}
\end{equation}
where $f_{\mathcal{G}}:\mathbb{R}^{2N} \rightarrow \mathbb{R}^{p|\mathcal{E}|}$ with $p$ depending on the formation control strategies.}
For the distance-based method, we have $p=1$, and the constraint \eqref{constraint} is
\begin{equation}
\begin{aligned}
f_{\mathcal{G}}^{\text {distance}}\left(x^{\text {end}}\right): & =\operatorname{col}_{(i, j) \in \mathcal{E}}\left(\ldots,\left\|x_i^{\text {end}}-x_j^{\text {end}}\right\|^2, \ldots\right) \\
& =f_{\mathcal{G}}^{\text {distance}}\left(x^*\right).
\end{aligned} \label{defi: edge function1}
\end{equation}
Regarding to this method, the use of 
rigidity graph framework as expounded in \cite{anderson2008rigid} plays an important role for the design and analysis of distributed formation controller in literature. The framework uses the notion of {\it infinitesimally rigid} formation, where roughly speaking, the formation shape is invariant under an infinitesimal motion of the agents. The framework $(\mathcal G,x^*)$ is called {\it infinitesimally rigid} if the rank of $ \displaystyle \frac{\partial f^\text{distance}_{\mathcal{G}}}{\partial x^{\text {end}}}(x^*)$ equals $2N-3$ (for 2D shape). Throughout this paper, whenever we discuss the distance-based method, we assume that the {\bb framework is {\it infinitesimally rigid}.} 
For the displacement-based method, we have $p=2$, and the constraint \eqref{constraint} is
\begin{equation}
\begin{aligned}
f_{\mathcal{G}}^{\text {displacement}}\left(x^{\text {end}}\right) :& =\operatorname{col}_{(i, j) \in \mathcal{E}}\left(\ldots, x_i^{\text {end}}-x_j^{\text {end}}, \ldots\right) \\
& = f_{\mathcal{G}}^{\text {displacement}}\left(x^*\right).
\end{aligned} \label{defi: edge function2}
\end{equation}
Interested readers can refer to \cite{Chan2021,Marina2016,mehdifar2020prescribed} and the included references for details about the rigid formation graph.

\section{Proposed Distributed Formation Controller} 
\label{main result}
\begin{figure}
	\centering
\includegraphics[scale=0.38]{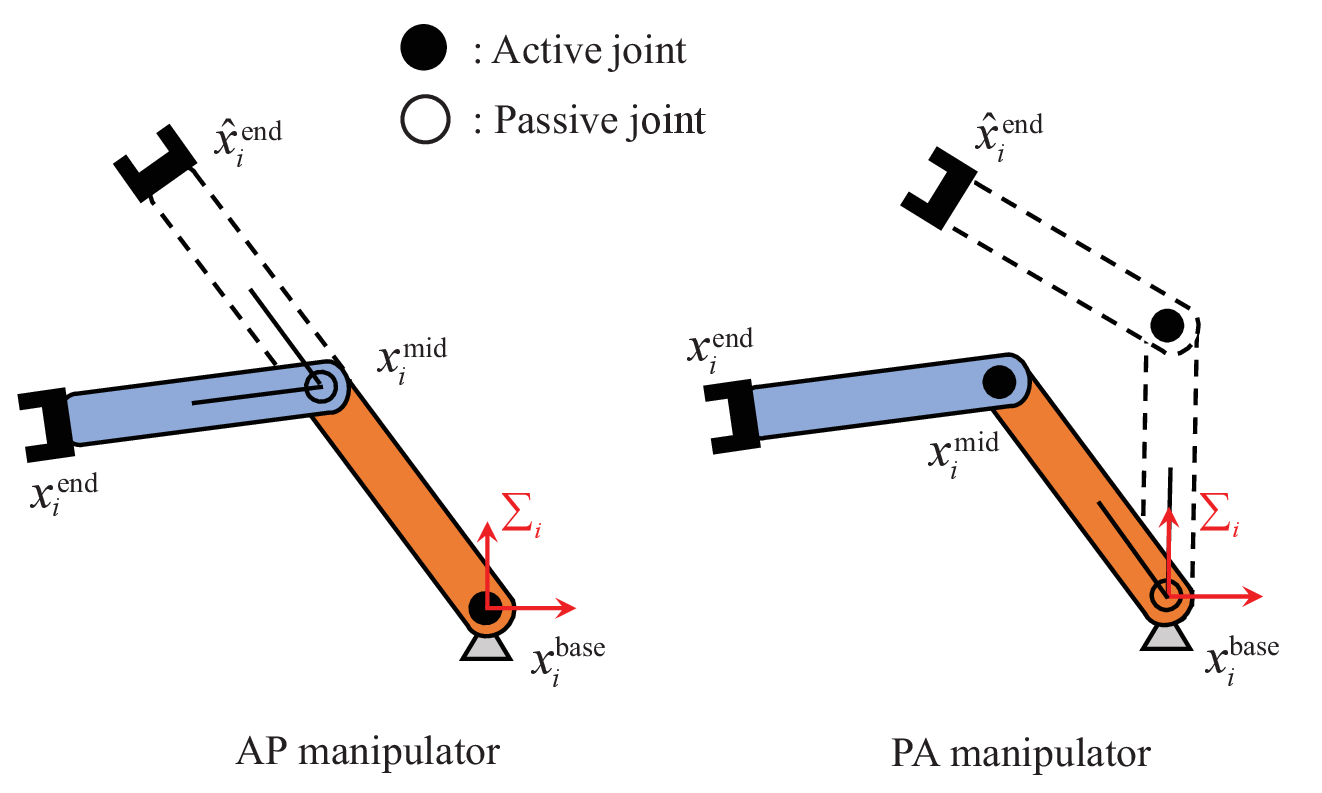}
	\caption{For the underactuated manipulator, the defined virtual end-effector position $\widehat x_i^\text{end}$ is established by setting the active joint angle to its current value and the passive joint angle to 0. {\bb For brevity, the torsional springs at the joints are omitted in the figure.}} 
 \label{fig: artificial position}
\end{figure}

\begin{figure*}
	\centering	\includegraphics[scale=0.55]{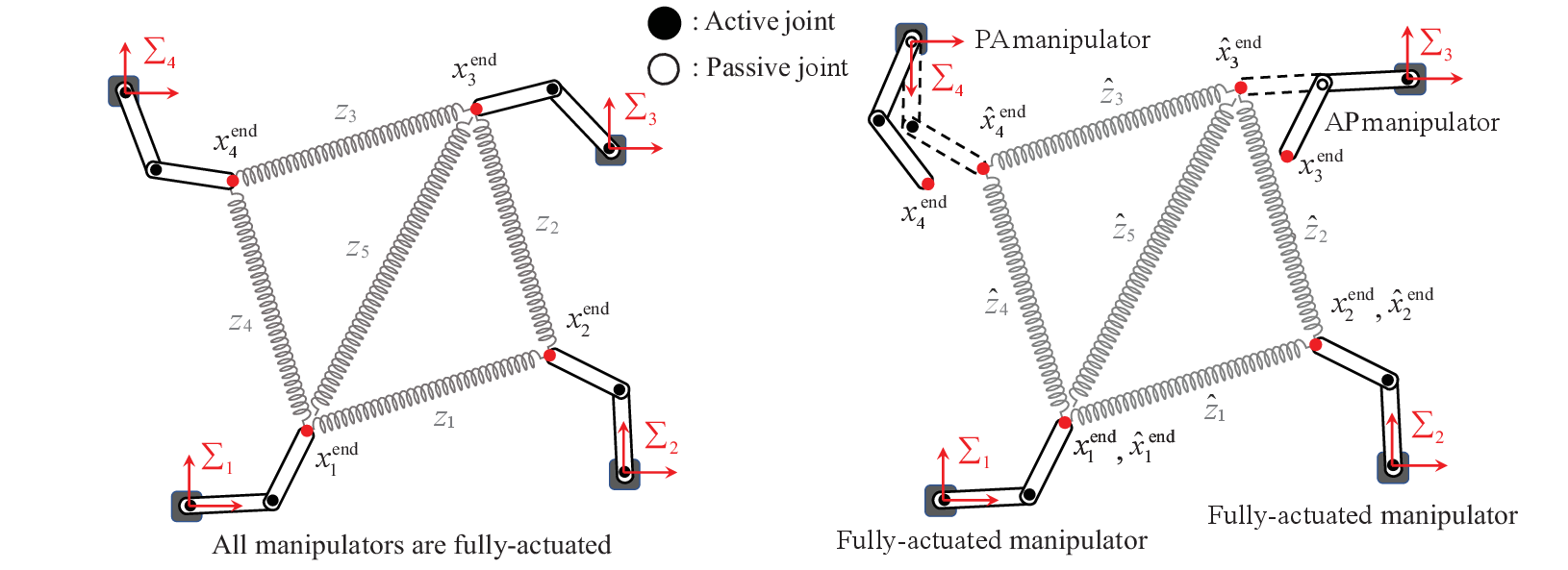}
	\caption{Distance-based method. In this case, the framework $(\mathcal{G},x^*)$ needs to be {\it infinitesimally rigid} and the springs in gray are the virtual coupling assigned to $\mathcal{G}$. (Left) In \cite{wu2021distributed,wu2022distributed}, all manipulators in the group are fully-actuated and the virtual springs are set to connect the end-effectors. (Right) In this paper, some manipulators are underactuated, and the virtual springs are set to connect the defined virtual end-effector positions. {\bb For brevity, the torsional springs at the joints are omitted in the figure.}} 
 \label{fig2}
\end{figure*}

\begin{figure*}
	\centering
	\includegraphics[scale=0.55]{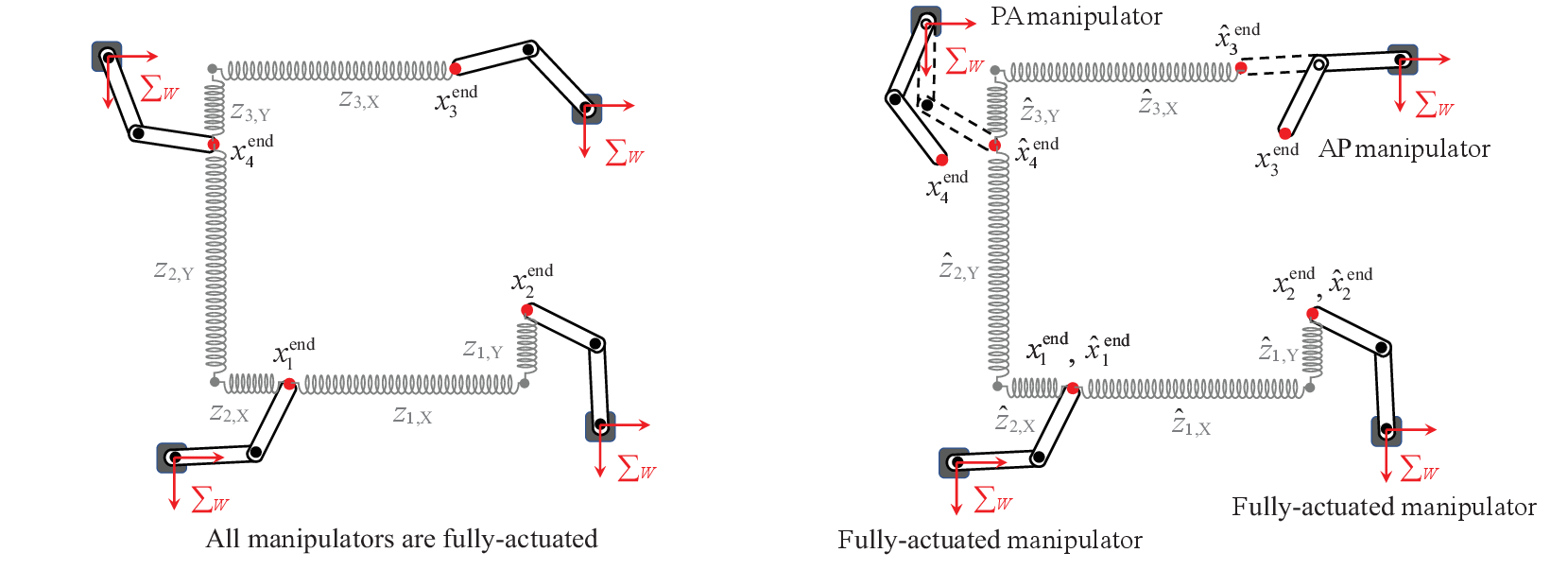}
	\caption{Displacement-based method. In this case, $\mathcal{G}$ needs to be connected and the springs in gray are the virtual coupling assigned to $\mathcal{G}$. (Left) In \cite{wu2021distributed,wu2022distributed}, all manipulators in the group are fully-actuated and the virtual springs are set to connect the end-effectors. (Right) In this paper, some manipulators are underactuated, and the virtual springs are set to connect the defined virtual end-effector positions. Note that the displacement-based method requires all manipulators to share the same coordinate frame \cite{wu2021distributed,wu2022distributed}. {\bb For brevity, the torsional springs at the joints are omitted in the figure.}} 
 \label{fig: displacement-based}
\end{figure*}

In this section, we will follow and modify the distance-based and displacement-based methods presented in \cite{wu2021distributed,wu2022distributed} to solve Problem \ref{problem}. For a group of fully-actuated manipulators, Wu et al. \cite{wu2021distributed,wu2022distributed} set virtual springs to connect the edges of the framework $(\mathcal G,x^\text{end})$, which means that every manipulator $i$ adjusts $x_i^\text{end}$ by actuating its joint angle $q_i$ in response to its neighboring end-effector positions $x_{j}^\text{end}$ for all ${j\in \mathcal N_i}$. 
When some of the manipulators are underactuated, the control strategy is no longer applicable, because there is no direct expression from active joint angle $q^\text{a}$ to the end-effector position $x^\text{end}$. Therefore, in Section \ref{sec:definition}, we define a stacked virtual end-effector position vector $\widehat x^\text{end} \in \mathbb{R}^{2N}:=\operatorname{col}_{i \in \mathcal{M}}\left(\ldots, \widehat x_i^\text{end}, \ldots \right)$, where $\widehat x_i^\text{end} \in \mathbb{R}^2$ is only related to the active joint angle(s) of the manipulator $i$ (Fig. \ref{fig: artificial position}). In Section \ref{sec:virtual spring}, virtual springs are defined to connect the edges of the framework $(\mathcal G,\widehat x^\text{end})$ to solve Problem \ref{problem} (Figs. \ref{fig2} and \ref{fig: displacement-based}). Particularly, due to the use of $\widehat{x}^{\text{end}}$, Problem \ref{problem} is adapted to
designing distributed controllers of the form \eqref{form} such that as $t \to \infty$, we have
\begin{description}
\item[{\bf 1.}]  $\widehat x^\text{end}(t) \to \mathcal{S}_W$; and
\item[{\bf 2.}]  $x^\text{end}(t) \to \widehat x^\text{end}(t)$ and $\dot q(t) \to {\bf 0}.$
\end{description}
Then, for the proposed control strategies, we discuss the cardinality of the set of desired and reachable shapes $\mathcal{S}_W$  for networked end-effectors in Section \ref{sec:work} and give a stability analysis for closed-loop systems in Section \ref{sec:stability}.

\subsection{Definition of the Virtual End-effector Position}
\label{sec:definition}
Let us define the virtual end-effector position $\widehat x_i ^\text{end} = {[{{\widehat x}_{i,X}^\text{end}},{{\widehat x}_{i,Y}^\text{end}}]^{\rm{T}}}$ for each type of manipulator $i$ as follows.

1) {\it The $i$-th manipulator is fully-actuated}: 
Let its virtual end-effector position be equal to the actual end-effector position, that is 
\begin{equation}
\widehat x_i^\text{end}=x_i^\text{end}, \; i \in \mathcal{M}_{\mathrm{fa}}, \label{eq:hatx_full}
\end{equation}
where $x_i^\text{end}$ is given in (\ref{task-space}). 

2) 
{\it The $i$-th manipulator is the AP manipulator}: As shown in the left sub-plot of Fig. \ref{fig: artificial position}, let its virtual end-effector position be
\begin{equation}
{{\widehat x}_i^\text{end}}: = {h_i}\left( {{q_{i,1}},0} \right) + {x_{i}^\text{base}}, \; i \in \mathcal{M}_{\mathrm{ap}},
\label{x1}
\end{equation}
where $h_i$ is given in \eqref{task-space}. From Fig. \ref{fig: artificial position}, by the geometrical relation, 
${{\widehat x}_i^\text{end}}$ can also be expressed using  ${x}_{i}^\text{mid}$ by 
\begin{equation}
\widehat{x}_i^\text{end}=\left(\frac{L_{i,1}+L_{i,2}}{L_{i,1}}\right)\left(x_{i}^\text{mid}-x_{i}^\text{base}\right)+x_{i}^\text{base}, \; i \in \mathcal{M}_{\mathrm{ap}}.\label{x1_anothoer}
\end{equation}
3) {\it The $i$-th manipulator is the PA manipulator}: As shown in the right sub-plot of Fig. \ref{fig: artificial position}, let its virtual end-effector position be 
\begin{equation}
{{\widehat x}_i^\text{end}}: = {h_i}\left( {0,{q_{i,2}}} \right) + {x_{i}^\text{base}}, \; i \in \mathcal{M}_{\mathrm{pa}}.
\label{x2}
\end{equation}
Define vectors $ {\bf a} = [a_1, a_2]^{\rm T}= {x_{i}^\text{mid}} - {x_{i}^\text{base}}$ and $ {\bf b} = [b_1, b_2]^{\rm T} = {x_{i}^\text{end}} - {x_{i}^\text{mid}}$, and note from Fig. \ref{fig1} that $q_{i,2}$ is the counterclockwise angle between ${\bf a}$ and ${\bf b}$. 
From a geometrical relation, it follows that 
\begin{equation}
\begin{aligned}
& q_{i, 2}=g\left(x_i^\text{end}, x_{i}^\text{mid}\right)+2 k \pi, \quad i \in \mathcal{M}_{\mathrm{pa}}, \\
& g\left(x_i^\text{end}, x_{i}^\text{mid}\right):=\operatorname{atan2}\;(\mathbf{a} \times \mathbf{b}, \mathbf{a} \cdot \mathbf{b}),
\end{aligned} \label{atan2}
\end{equation}
where $\operatorname{atan2}$ is the two-argument arctangent function \cite[Appendix A.1]{spong2006robot} and $k \in \mathbb{Z}$. Here, the symbol $\times$ denotes the pseudo-cross product and ${\bf{a}} \times {\bf{b}} = \left| {\bf{a}} \right|\left| {\bf{b}} \right|\sin {q_{i,2}} = {a_1}{b_2} - {a_2}{b_1}$; the notation $ \cdot $ denotes the dot product and $ {\bf a} \cdot {\bf b} = \left| {\bf{a}} \right|\left| {\bf{b}} \right|\cos {q_{i,2}}= {a_1}{b_1} + {a_2}{b_2}$. Then, we can rewrite (\ref{x2}) as
\begin{equation}
{{\widehat x}_i^\text{end}} = {h_i}\left( 0,g({x_i^\text{end}},{x_{i}^\text{mid}})\right) + {x_{i}^\text{base}}, \; i \in \mathcal{M}_{\mathrm{pa}}.
\label{x2_anothoer}
\end{equation}
Correspondingly, differentiating (\ref{eq:hatx_full}), (\ref{x1}) and (\ref{x2}) gives
\begin{equation}
   \dot{\widehat{x}}_i^{\text{\raisebox{-0.8ex}{end}}}= \begin{cases}J_i\left(q_i\right) \dot{q}_i, & i \in \mathcal{M}_{\mathrm{fa}} , \vspace{0.2cm} \\ r_i \bar{J}_i\left(q_{i, 1}\right) \dot{q}_{i, 1}, & i \in \mathcal{M}_{\mathrm{ap}}, \vspace{0.2cm} \\ r_i \bar{J}_i\left(q_{i, 2}\right) \dot{q}_{i, 2}, & i \in \mathcal{M}_{\mathrm{pa}},\end{cases}
   \label{defi:dot xi}
\end{equation}
where $J_i(q_i) \in \mathbb{R}^{2\times2}$ is defined in (\ref{jaco}) and $\bar{J}_i(\cdot):=\left[\begin{array}{ll}
\bar{J}_{i, 1}(\cdot) & \bar{J}_{i, 2}(\cdot)
\end{array}\right]^{\mathrm{T}}=\left[-\cos \left(\cdot+\beta_i\right)-\sin \left(\cdot+\beta_i\right)\right]^{\mathrm{T}}$.
The positive constant $r_i$ is associated with the length of the manipulator's links, which is defined as $r_i = L_{i, 1}+L_{i, 2}$ for $i \in \mathcal{M}_{\mathrm{ap}}$ and as $r_i = L_{i, 2}$ for $i \in \mathcal{M}_{\mathrm{pa}}$.

Now, we present the following properties about the equilibrium configuration of the studied underactuated manipulators, which are crucial for subsequent controller design and stability analysis. These properties guarantee that $\widehat{x}_i^\text{end} = x_i^\text{end}$ for each underactuated manipulator $i \in \mathcal{M}_{\mathrm{ap}} \cup \mathcal{M}_{\mathrm{pa}}$ when it reaches the steady state.

\begin{lemma}
Consider the AP manipulator $i \in \mathcal{M}_{\mathrm{ap}}$ as in (\ref{full}). If its active joint angle
$q_{i,1}(t)$ is constant for all $t\geq 0$ under a constant torque $u_{i,1}$ and the passive joint angular velocity $\dot q_{i,2}(t)$ is bounded, then $q_{i,2}(t) = 0$ for all $t\geq 0$. \label{lemma_AP}
\end{lemma}

\begin{lemma}
Consider the PA manipulator $i \in \mathcal{M}_{\mathrm{pa}}$ as in (\ref{full}) with its mechanical parameters satisfying $\alpha_{i,2} \ne \alpha_{i,3}$. If the active joint angle
$q_{i,2}(t)$ is constant for all $t\geq 0$ under a constant torque $u_{i,2}$, then $q_{i,1}(t) = 0$ for all $t\geq 0$. 
\label{lemma_PA}
\end{lemma}

The proofs of Lemmas \ref{lemma_AP} and \ref{lemma_PA} are given respectively in Appendices \ref{app A} and \ref{app B}.  

\subsection{The Virtual Spring Setting for Formation Control}
\label{sec:virtual spring}
Let's denote the relative displacement between the actual end-effector positions as $z$ and that between the virtual end-effector positions as $\widehat z$
\begin{equation}
\begin{aligned}
& z:=\operatorname{col}_{k \in\{1, \ldots, \mid \mathcal{E}\mid\}}\left(\ldots, z_k, \ldots\right)=\bar{B}^{\mathrm{T}} x^{\mathrm{end}}, \\
& \widehat{z}:=\operatorname{col}_{k \in\{1, \ldots, \mid \mathcal{E}\mid\}}\left(\ldots, \widehat{z}_k, \ldots\right)=\bar{B}^{\mathrm{T}} \widehat{x}^{\mathrm{end}}, \\
& z_k:=x_i^{\text {end}}-x_j^{\text {end}}, \;\; \widehat{z}_k:=\widehat{x}_i^{\text {end}}-\widehat{x}_j^{\text {end}},
\end{aligned} 
\end{equation}
where $\bar B \in \mathbb{R}^{2N\times2|\mathcal{E}|} := B\otimes I_2$ and $(i, j)=\mathcal{E}_k \in \mathcal{E}$.

Unlike \cite{wu2022distributed} and \cite{wu2021distributed}, we use virtual springs to connect the virtual end-effector position $\widehat x^\text{end}$ instead of the actual position $x^\text{end}$, as illustrated in Figs. \ref{fig2} and \ref{fig: displacement-based}. 
The virtual springs are considered to be in their natural states and reach the minimum potential energy when the desired formation is achieved. For edge $\mathcal{E}$ of the framework $(\mathcal G,\widehat x^\text{end})$, denote the error signal
\begin{equation}
e=f_e(\widehat z):={f_\mathcal{G}}\left( {{\widehat x^{{\text{end}}}}} \right) - {f_\mathcal{G}}\left( {{x^*}} \right), \;  e\in \mathbb{R}^{p|\mathcal{E}|}. \label{eq:ek} 
\end{equation}
For the distance-based method, $f_\mathcal{G}$ is defined as \eqref{defi: edge function1} with $p=1$; while for the
displacement-based method, $f_\mathcal{G}$ is defined as \eqref{defi: edge function2} with $p=2$. 
The potential function $V (e)$ for the desired formation is
\begin{equation}
V(e): = \frac{1}{2}{e^T}{k_S}e, \label{Ve}
\end{equation}
where $k_S\in \mathcal{R}^{p|\mathcal{E}|\times p|\mathcal{E}|}$ is a diagonal matrix, whose diagonal entries are positive and represent the stiffness of the virtual springs.
Denote $\displaystyle D\left(\widehat z\right):=\frac{\partial e}{\partial \widehat z}\in \mathbb{R}^{2|\mathcal{E}|\times p|\mathcal{E}|}$, and routine computation shows that 
 \begin{equation}
 \widehat{e}:=\frac{\partial V(e)}{\partial \widehat x^\text{end}} = \bar BD(\widehat z) k_S e,   \label{defined}
\end{equation}
where 
\begin{equation}
\widehat{e}:=\operatorname{col}_{i \in\mathcal{M}}\left(\ldots, \widehat{e}_i, \ldots\right) \in \mathbb{R}^{2 N}, \text { with } \widehat{e}_i:=\frac{\partial V(e)}{\partial \widehat{x}_i^{\text {end}}}.
\label{hat e1}
\end{equation}
Note that ${D}({\widehat z}) =\operatorname{block~diag}_{k \in\{1, \ldots, |\mathcal{E}|\}}\left(\ldots, 2\widehat z_k, \ldots\right)$ \vspace{0.1cm} for the distance-based method, while ${D}({\widehat z}) = I_{2|\mathcal{E}|}$ for the displacement-based method. As discussed in \cite{Marina2016,wu2022distributed}, the matrix $D^\mathrm{T}(\widehat z)\bar B^\mathrm{T} \bar BD(\widehat z)$ with $f_e(\widehat z)={\bf 0}$ (or when $\widehat{x}^{\text{end}}$ forms the desired shape) is positive definite if the framework $(\mathcal{G},{x}^*)$ is {\it infinitesimally rigid}; and the matrix $ B^\mathrm{T}  B$ is positive definite if $\mathcal{G}$ includes no cycles. 

\subsection{The Cardinality of \texorpdfstring{$\mathcal{S}_W$}{\mathcal{S}_W}}
\begin{figure}
	\centering
	\includegraphics[scale=0.55]{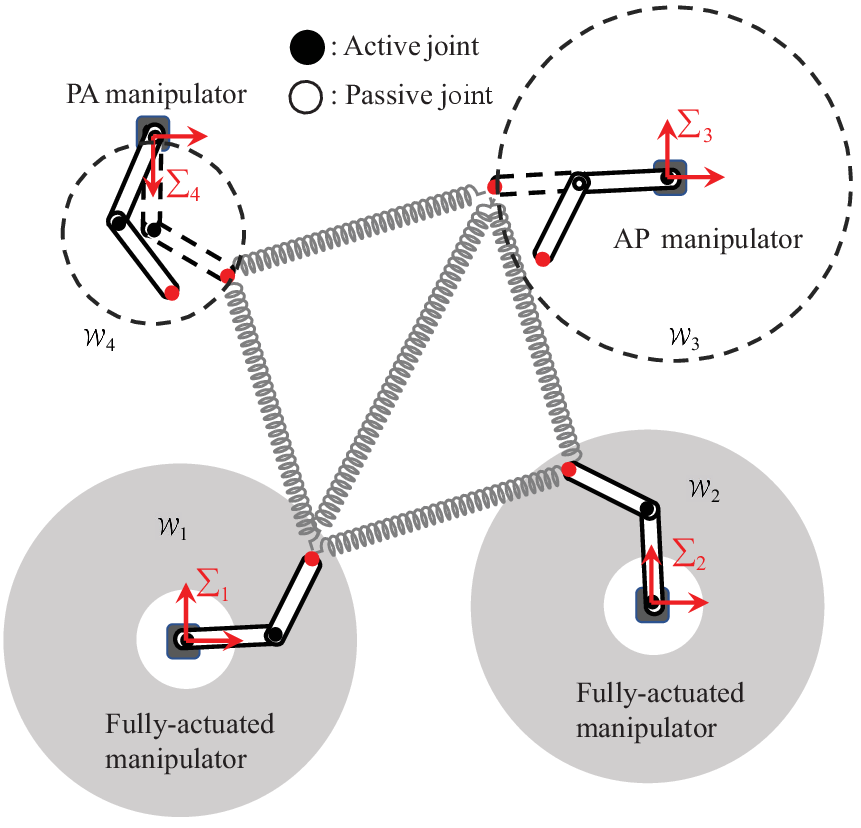}
	\caption{The reachable space of the networked manipulators. For the fully-actuated manipulator $i$, its reachable space $\mathcal{W}_i$ is a subset of a disk, represented by the gray area; for the underactuated manipulator $i$, its reachable space is a circle, represented by the black dashed line. {\bb For brevity, the torsional springs at the joints are omitted in the figure.}}  
 \label{fig4}
\end{figure}
\label{sec:work}
Before the distributed formation controller design, we need to ascertain whether a given desired formation shape is achievable by the networked manipulators, i.e., if the set of desired and reachable shapes $\mathcal{S}_W$ is non-empty. From Lemmas \ref{lemma_AP} and \ref{lemma_PA}, we know that $x^\text{end} = \widehat x^\text{end}$ when the manipulators are in steady state. Therefore, the reachable space $\mathcal{W}_i$ of the manipulator $i$, as visually depicted in Fig. \ref{fig4}, is defined as follows, where the base position is denoted by $x_{i}^\text{base}$.

1) For the fully-actuated manipulator $i \in \mathcal{M}_{\mathrm{fa}}$, 
we define the reachable space $\mathcal W_i$ by
\begin{equation}
{\mathcal{W}_i:=\left\{\widehat{x}_i^\text{end}: \widehat{x}_i^\text{end}=h_i\left(q_{i}\right)+x_{i}^\text{base}, \; q_{i}\in \rline^2\right\}.} \label{W_fa}
\end{equation}

2) For the AP manipulator $i \in \mathcal{M}_{\mathrm{ap}}$, 
its reachable space is 
\begin{equation}
 \mathcal{W}_i:=\left\{\widehat{x}_i^\text{end}: \widehat{x}_i^\text{end}=h_i\left(q_{i, 1}, 0\right)+x_{i}^\text{base},\;q_{i,1}\in \rline \right\},  \label{W_ap}
\end{equation}
which is parameterized by the active joint angle $q_{i,1}$. Note that it is a circle with its center at $x_{i}^\text{base}$ and a radius of $r_i = L_{i,1}+L_{i,2}$. 

3) For the PA manipulator $i \in \mathcal{M}_{\mathrm{pa}}$, 
its reachable space is 
\begin{equation}
\mathcal{W}_i:=\left\{\widehat{x}_i^\text{end}: \widehat{x}_i^\text{end}=h_i\left(0, q_{i, 2}\right)+x_{i}^\text{base},\;q_{i,2}\in \rline\right\},  \label{W_pa}
\end{equation}
which is parameterized by the active joint angle $q_{i,2}$. In this case, $\mathcal W_i$ is a circle centered at $x_{i}^\text{base}+\left[L_{i, 1} \sin \left(\beta_i\right) \; L_{i, 1} \cos \left(\beta_i\right)\right]^{\rm T}$ and with a radius of $r_i = L_{i,2}$. 

Hence, the reachable space of the networked manipulators $\mathcal{W}:=\mathcal{W}_1 \times \cdots \times \mathcal{W}_N$ is parameterized by the stacked vector $q^\text{a} \in \mathbb R^{N+N_1}$ defined in (\ref{qa}), which means that the degree of freedom of $\widehat x^\text{end} \in \mathcal{W}$ is $N+N_1$. Now, we are ready to discuss the 
cardinality of $\mathcal{S}_W$.

For a given reference end-effector position $x^*$, if the framework ($\mathcal{G}$, $x^*$) is {\it infinitesimally rigid}, based on \eqref{defi: edge function1}, we can use the set 
\begin{equation}
\mathcal{S}^{{\text{distance}}}_{W}:= \left\{ {{{\widehat x}^{{\text{end}}}}\in \mathcal{W}:f_\mathcal{G}^{{\text{distance}}}({\widehat x^{{\text{end}}}}) = f_\mathcal{G}^{{\text{distance}}}({x^*})} \right\}
\label{z_distance}
\end{equation}
to locally define the set of desired and reachable shapes in distance-based control.
According to \cite{asimow1979rigidity}, if the framework is {\it infinitesimally rigid}, then the graph $\mathcal{G}$ contains exactly $2N-3$ independent edges (for 2D shape), which means that the set \eqref{z_distance} is determined by $2N-3$ independent defining equations. By comparing the degree of freedom $N+N_1$ of $\widehat x^\text{end}$ and the number of independent defining equations $2N-3$, we can deduce the cardinality of $\mathcal{S}^\text{distance}_{W}$, as discussed in the following remark. 

\begin{remark}
For a group of manipulators with the given suitable $x_i^\text{base}$ and $\beta_i, i=1,...,N$, there is
an inverse correlation between the cardinality of $\mathcal{S}_W^\text{distance}$ and the count of underactuated manipulators $N_2+N_3$ within the group. Specifically, if $N_2+N_3 \le 2$, the set $\mathcal{S}_W^\text{distance}$ is defined by an underdetermined system of equations and possesses an infinite cardinality; otherwise if $ N_2+N_3 = 3$, the set $\mathcal{S}_W^\text{distance}$ is defined by a well-determined system of equations and has a finite positive cardinality; else if $ N_2+N_3 \ge 4$,  the set $\mathcal{S}_W^\text{distance}$ is defined by an overdetermined system of equations and can be an empty set. 
\label{re:sw1}
\end{remark}

If $\mathcal{G}$ is connected, based on \eqref{defi: edge function2}, we can use the set
\begin{equation}
\resizebox{\columnwidth}{!}{
$\mathcal{S}^{{\text{displacement}}}_{W}:= \left\{ {{{\widehat x}^{{\text{end}}}}\in \mathcal{W}:f_{\mathcal{G}}^{{\text{displacement}}}({\widehat x^{{\text{end}}}}) = f_\mathcal{G}^{{\text{displacement}}}({x^*})} \right\}$
}
\label{z_displacement}
\end{equation}
to uniquely define the set of desired and reachable shapes in displacement-based control. Note that ${\mathcal{S}}^{{\text{displacement}}}_W\subset {\mathcal{S}}^{{\text{distance}}}_W$, since the displacement-based method leads to $R = I_2$ in (\ref{S}), i.e., it only accepts the desired formation shapes obtained by translation to $x^*$. Since $\mathcal{G}$ is connected, it contains exactly $N-1$ independent edges, 
which implies that the set \eqref{z_displacement} is determined by $2N-2$ independent defining equations.\vspace{0.1cm} Similar as before, we can deduce the cardinality of $\mathcal{S}^\text{displacement}_{W}$, as stated below. 

\begin{remark}
For a group of manipulators with the given suitable $x_i^\text{base}$ and $\beta_i, i=1,...,N$, the cardinality of $\mathcal{S}_W^\text{displacement}$ is less than that of $\mathcal{S}_W^\text{distance}$. Specifically, if $N_2+N_3 \le 1$ then $\mathcal{S}_W^\text{displacement}$ has an infinite cardinality; otherwise if $ N_2+N_3 = 2$ then $\mathcal{S}_W^\text{displacement}$ has a finite positive cardinality; else if $ N_2+N_3 \ge 3$ then $\mathcal{S}_W^\text{displacement}$ can be an empty set. 
\label{re:sw2}
\end{remark}

\subsection{Controller Design and Stability Analysis}
\label{sec:stability}

Prior to presenting distributed formation control laws, we 
need the following assumption.

\begin{assumption}
The set of desired and reachable shapes $\mathcal S_W$ is not empty and the networked manipulators avoid the set of
singular points as given below. Specifically, there exists a neighborhood $\mathcal S_{W_\mu}$ of $\mathcal S_W$  
such that {for all $q^\text{a}$ satisfying $\widehat x^{\rm end} (q^\text{a})\in \mathcal S_{W_\mu}$}, we have
\begin{description}
\item[{\bf A1.}] $q_{i, 2}+\beta_i \neq k \pi \; (k \in \mathbb{Z})$ for the fully-actuated manipulator $ i \in \mathcal{M}_{\mathrm{fa}}$; 
\item[{\bf A2.}] $\displaystyle q_{i, 1}+\beta_i \neq \frac{k \pi}{2}(k \in \mathbb{Z})$ for the AP manipulator $i \in \mathcal{M}_{\mathrm{ap}}$; and 
\item[{\bf A3.}] $\displaystyle q_{i, 2}+\beta_i \neq \frac{k \pi}{2}(k \in \mathbb{Z})$ for the PA manipulator $i \in \mathcal{M}_{\mathrm{pa}}$. 
\end{description} \label{Ass:3}
Here, the neighborhood $\mathcal{S}_{W_\mu}:=\left\{\widehat x^{\rm end} \in \mathbb{R}^{2 N}:\|e\| \le \mu \right\}$ with $\mu$ being some positive constant.
\end{assumption}

From Remarks \ref{re:sw1} and \ref{re:sw2}, we know that for networked manipulators with suitable fixed $x_i^\text{base}, \beta_i, i=1,...,N$, if the number of underactuated manipulators is not more than 3 for the distance-based method (or 2 for the displacement-based method), then $\mathcal{S}_W$ is not empty. On the other hand, the continuity argument guarantees that {\bf A1}--{\bf A3} are
met when choosing the desired shape at a reference point $x^*$ in $\mathcal{S}_W$ where $q_i=q_{i}^\text{ref}$, such that 
\[\begin{cases}  q_{i, 2}^\text{ref}+\beta_i \neq k \pi \; (k \in \mathbb{Z}), &\text {for all } i \in \mathcal{M}_{\mathrm{fa}}, \vspace{0.1cm}\\ \displaystyle q_{i, 1}^\text{ref}+\beta_i \neq \frac{k \pi}{2} \; (k \in \mathbb{Z}), &\text {for all } i \in \mathcal{M}_{\mathrm{ap}}, \vspace{0.1cm}\\ \displaystyle q_{i, 2}^\text{ref}+\beta_i \neq \frac{k \pi}{2}\;(k \in \mathbb{Z}), &\text {for all } i \in \mathcal{M}_{\mathrm{pa}}.\end{cases}\]

\begin{theorem}
	\label{the}
Consider the end-effector formation
control problem of mixed planar fully- and under-actuated
manipulators with flexible joints stated in Problem \ref{problem}. Under Assumption \ref{Ass:3}, the following
distributed controllers can be used to solve the problem locally
\begin{equation}
 u_i= \begin{cases}- J_i^{\mathrm{T}}\left(q_i\right) \widehat{e}_i-k_{D} \dot{q}_i+K_i q_i,\qquad \qquad \quad i \in \mathcal{M}_{\mathrm{fa}}, \vspace{0.2cm}\\
{\left[\begin{array}{c}
-r_i \bar{J}_i^{\rm\;T}\left(q_{i, 1}\right) \widehat{e}_i-k_{D} \dot{q}_{i, 1}+K_{i, 1} q_{i, 1} \\
0
\end{array}\right], i \in \mathcal{M}_{\mathrm{ap}}}, \vspace{0.2cm}\\
{\left[\begin{array}{c}
0 \\
 -r_i \bar{J}_i^{\rm\;T}\left(q_{i, 2}\right) \widehat{e}_i-k_{D} \dot{q}_{i, 2}+K_{i, 2} q_{i, 2}
\end{array}\right], i \in \mathcal{M}_{\mathrm{pa}}},\end{cases} 
 \label{eq:controller}
\end{equation}	
where $J_i(q_i) \in \mathbb{R}^{2 \times 2}$, $\bar J_i(\cdot) \in \mathbb{R}^{2}$ and the positive constant $r_i$ are given in \eqref{defi:dot xi}; the vector $\widehat e_i \in \mathbb{R}^{2}$ is defined in \eqref{hat e1}; the positive constant $k_{D}$ is the controller gain.
\end{theorem}

\begin{remark}
Please note that the distributed controllers (\ref{eq:controller}) align with the form illustrated in (\ref{form}). The term $\widehat e_i$ within (\ref{eq:controller}) is a function that depends only on $\widehat z_k: = \widehat x_i^\text{end}-\widehat x_j^\text{end}, j \in \mathcal N_i$. 
Notice that $\widehat x_i^\text{end}$ can be expressed by $q_i$, as shown by (\ref{eq:hatx_full}), (\ref{x1}), and (\ref{x2}), while $\widehat x_j^\text{end}$ can be expressed by $x_j^\text{end}$ and $x_{j}^\text{mid}$, as shown by (\ref{eq:hatx_full}), (\ref{x1_anothoer}), and (\ref{x2_anothoer}).
\end{remark}

\begin{proof}
Now, we give a proof of Theorem \ref{the}. We can rewrite 
\eqref{eq:controller}
into a compact form as follows
\begin{equation}
\bar{u}=-J^{\mathrm{T}}(q^\mathrm{a}) \widehat{e}-{k_D} \dot{q}^\mathrm{a}+ K^\mathrm{a} q^\mathrm{a}, \label{controller}
\end{equation}
where 
the vector $q^\mathrm{a}\in \mathbb{R}^{N+N_1}$ and the matrix $K^\mathrm{a} \in \mathbb{R}^{({N+N_1}) \times ({N+N_1})}$ are defined in (\ref{qa}) and (\ref{defi:ka}), respectively; the stacked vector $\widehat e \in \mathbb{R}^{2N}$ is given in (\ref{defined}); the stacked vector of all active control torques $\bar u\in \mathbb{R}^{N+N_1}$ and the matrix $J\left(q^\mathrm{a}\right)\in \mathbb{R}^{2N \times (N+N_1)}$ are 
\[\begin{aligned}
& \bar{u}:=\operatorname{col}\big\{\operatorname{col}_{i \in \mathcal{M}_{\mathrm{fa}}}\left(\ldots, u_i, \ldots\right), \operatorname{col}_{i \in \mathcal{M}_{\mathrm{ap}}}\left(\ldots, u_{i, 1}, \ldots\right),\\
&\qquad\qquad\left.\operatorname{col}_{i \in \mathcal{M}_{\mathrm{pa}}}\left(\ldots, u_{i, 2}, \ldots\right)\right\}, \\
& J\left(q^{\mathrm{a}}\right):=\operatorname{block\;diag} \left(J_{\mathrm{fa}}\left(q_{\mathrm{fa}}\right), \bar{J}_{\mathrm{ap}}\left(q_{\mathrm{ap}}^{\mathrm{a}}\right), \bar{J}_{\mathrm{pa}}\left(q_{\mathrm{pa}}^{\mathrm{a}}\right)\right),
\end{aligned}
\]
with $J_{\mathrm{fa}}\left(q_{\mathrm{fa}}\right) :=\operatorname{block\;diag}_{i \in \mathcal{M}_{\mathrm{fa}}}\left(\ldots, J_i\left(q_i\right), \ldots\right)$,  $\bar J_{\mathrm{ap}}\left(q_{\mathrm{ap}}^{\mathrm{a}}\right):=\operatorname{block\;diag}_{i \in \mathcal{M}_{\mathrm{ap}}}\left(\ldots, r_i \bar{J}_i\left(q_{i, 1}\right), \ldots\right)$, and $\bar J_{\mathrm{pa}}\left(q_{\mathrm{pa}}^{\mathrm{a}}\right):=\operatorname{block\;diag}_{i \in \mathcal{M}_{\mathrm{pa}}}\left(\ldots, r_i \bar{J}_i\left(q_{i, 2}\right), \ldots\right)$.


Consider the  Lyapunov function
\begin{equation}
U = V(e) + \frac{1}{2}{{\dot q}^{\rm{T}}}M(q)\dot q + \frac{1}{2}{({q^\text{u}})^{\rm{T}}}{K^\text{u}}{q^\text{u}},\label{U}
\end{equation}
where $V(e)$ is defined in (\ref{Ve}) and $q^\text{u} \in \mathbb{R}^{N_2+N_3}, K^\text{u} \in \mathbb{R}^{(N_2+N_3) \times (N_2+N_3)}$ are defined in (\ref{qa}) and (\ref{defi:ka}), respectively. A routine computation to the time derivative of (\ref{U}) yields
\begin{align}
\begin{aligned}
\dot{U}= & \left(\frac{\partial V}{\partial \widehat{x}^\text{end}}\right)^{\mathrm{T}}  \dot{\widehat{x}}^{\smash{\raisebox{-0.8ex}{\hspace{0em}$\scriptstyle\text{end}$}}}+\dot{q}^{\mathrm{T}}(M(q) \ddot{q}) \\
& +\frac{1}{2} \dot{q}^{\mathrm{T}} \dot{M}(q) \dot{q}+\left(\dot{q}^\text{u}\right)^{\mathrm{T}} K^\text{u} q^\text{u} \\
= & \; \widehat{e}^{\mathrm{T}} J\left(q^\text{a}\right) \dot{q}^\text{a}+\dot{q}^{\mathrm{T}}(u-C(q, \dot{q}) \dot{q}-K q) \\
& +\frac{1}{2} \dot{q}^{\mathrm{T}} \dot{M}(q) \dot{q}+\left(\dot{q}^\text{u}\right)^{\mathrm{T}} K^\text{u} q^\text{u} \\
= & \;\widehat{e}^{\mathrm{T}} J\left(q^\text{a}\right) \dot{q}^\text{a}+\dot{q}^{\mathrm{T}} u-\dot{q}^{\mathrm{T}} K q+\left(\dot{q}^\text{u}\right)^{\mathrm{T}} K^\text{u} q^\text{u} \\
= & \;\widehat{e}^{\mathrm{T}} J\left(q^\text{a}\right) \dot{q}^\text{a}+\left(\dot{q}^\text{a}\right)^{\mathrm{T}} \bar{u}-\dot{q}^{\mathrm{T}} K q+\left(\dot{q}^\text{u}\right)^{\mathrm{T}} K^\text{u} q^\text{u}.
\end{aligned}
\label{dot U}
\end{align}
where the second equality arises from \eqref{defined}, \eqref{defi:dot xi} and the dynamics \eqref{dd}, while the third one arises from the fact that the matrix $\dot M_i(q_i) - 2C_i(q_i, \dot{q}_i)$ is skew-symmetric. Substituting the controller (\ref{controller}) into {\bb (\ref{dot U})} yields
\begin{equation}
\begin{aligned}
\begin{aligned}
\dot{U} & =-\left(\dot{q}^\text{a}\right)^{\mathrm{T}} k_D \dot{q}^\text{a}+\left(\dot{q}^\text{a}\right)^{\mathrm{T}} K^\text{a} q^\text{a}-\dot{q}^{\mathrm{T}} K q+\left(\dot{q}^\text{u}\right)^{\mathrm{T}} K^\text{u} q^\text{u} \\
& =- k_D \|\dot{q}^\text{a}\|^2.
\end{aligned}
\end{aligned} \label{dot U2}
\end{equation}
{\tb Therefore, we can deduce $V(e)=\displaystyle \frac{1}{2} k_S \|e\|^2 \le U(0)$. Consider that the
closed-loop systems initiate from a state ($q(0)$, $\dot q(0)$) near the desired formation shape with 
$\displaystyle U(0) \le \frac{1}{2} k_S \mu^2$. Then, we have $\|e(t)\| \le \mu$ so that $\widehat{x}^\text{end}(t)$ 
remains within $\mathcal{S}_{W_\mu}$ for all $t \geq 0$. }

{\tb Drawing on the stability analysis in \cite{fantoni2000energy,xin2023nonlinear}, we redefine a state for the closed-loop systems consisting of \eqref{dd} and \eqref{controller} facilitating the application of La-Salle’s invariance principle \cite[Theorem 4.4]{Kha02}}
\begin{equation}
    y=\operatorname{col}(C_{qa}, S_{qa}, q^{\rm u}, \dot q^{\rm a}, \dot q^{\rm u}).
\end{equation}
{\tb Here, \(C_{qa}\) and \(S_{qa}\) are vectors in \(\mathbb{R}^{N+N_1}\), whose elements are the cosine and sine, respectively, of the corresponding elements of the vector \(q^a\). Then, we express the closed-loop systems as}
\begin{equation}
\dot{y}=F(y)=\operatorname{col}\left(-S_{qa} \odot \dot{q}^{\mathrm{a}}, C_{qa} \odot \dot{q}^{\mathrm{a}}, \dot{q}^{\mathrm{u}}, \ddot{q}^{\mathrm{a}}, \ddot{q}^{\mathrm{u}}\right),
\end{equation}
{\tb where $\odot$ represents the Hadamard product, i.e., the element-wise multiplication of two vectors with the same dimensions. Note that both $\ddot{q}^{\text{a}}$ and $\ddot{q}^{\text{u}}$ are continuously differentiable functions of $y$, which arises from two key factors. Firstly, the angular variables present in $M(q)$, $C(\dot{q}, q)$, $J(q^{\rm a})$, $\widehat z$ and $e$ exclusively take trigonometric forms. Secondly, the linear terms of active joint angles $K^{a}q^{a}$ in \eqref{dd}, attributed to the torsional springs, are negated by the corresponding term in the controller \eqref{controller}.

Define a set $\Xi:=\{y : U \leq \gamma\}$ with $\gamma$ being a positive constant. Using the above state variable $y$, we will establish in the following paragraphs that as $t \to \infty$, every $y(t)$ starting in $\Xi$ converges to the equilibrium set}
\begin{equation}
\Omega := \{ y: {q^\text{u}} = {\bf 0}, \;{\dot q} = {\bf 0}, \; e(C_{qa},S_{qa}) = {\bf 0}\},  \label{eq:invar}
\end{equation}
{\tb and the control objective is obtained.}

{\tb From \eqref{U} and (\ref{dot U2}), we know $U$ is a continuously differentiable function of $y$ and remains bounded, which implies that $\dot q$, $q^\text{u}$ are bounded. Consequently, $y$ is bounded and the set $\Xi$ is compact and positively invariant. 
According to the LaSalle’s invariance principle \cite[Theorem 4.4]{Kha02}, as $t \to \infty$, the solution $y(t)$ of the closed-loop systems converges to the largest invariant set in }
\begin{equation}
\{ y:\dot U = 0\}  = \{ y:{\dot q^\text{a}} = {\bf 0}\}.  \label{eq:dot V=0}
\end{equation}
Substituting $\dot q^\text{a} = {\bf 0}$ into (\ref{defi:dot xi}) results in $\widehat x^\text{end}$ being constant vector, which further implies that $ \widehat e$ and $\bar u$ are constant vectors. Therefore, according to Lemmas \ref{lemma_AP} and \ref{lemma_PA}, under the assumption $\alpha_{i,2} \ne \alpha_{i,2}$ for all $i\in \mathcal{M}_\text{pa}$, we have $\dot q^\text{u} =  q^\text{u} = {\bf 0}$. Hence, we have $\dot q={\bf 0}$ and ${x^{{\rm{end }}}} = {{\widehat x}^{{\rm{end }}}}$. Consequently, from \eqref{dd}, we have $\bar u-K^\text{a}q^\text{a} = {\bf 0}$. Therefore, \eqref{controller} can be reduced to 
\begin{equation}
 - {J^\mathrm{T}( q^\text{a}) }\widehat e = {\bf 0}. \label{eq mother}
\end{equation}
We can decompose (\ref{eq mother}) into
\begin{align}
&\begin{aligned}
- J_i^\mathrm{T}(q_i) {{\widehat e}_i} = {\bf 0}, \quad \;\; i \in \mathcal{M}_{\mathrm{fa}}, \label{eq1}
\end{aligned}\\
&\begin{aligned}
- \bar J_i^\mathrm{\;T}(q_{i,1}) {{\widehat e}_i} = 0, \quad i \in \mathcal{M}_{\mathrm{ap}}, \label{eq2}
\end{aligned}\\
&\begin{aligned}
- \bar J_i^\mathrm{\;T}(q_{i,2}) {{\widehat e}_i} = 0, \quad i \in \mathcal{M}_{\mathrm{pa}}.\label{eq3}
\end{aligned} 
\end{align}
For all $i \in \mathcal{M}_{\mathrm{fa}}$, $J_i(q_i) \in \mathbb{R}^{2\times 2}$ is invertible if and only if $q_{i, 2}+\beta_i \neq k \pi \; (k \in \mathbb{Z})$. Since the closed-loop systems remain within $\mathcal{S}_{W_\mu}$ for all $t \geq 0$,
under {\bf A1} in Assumption \ref{Ass:3}, we have $ {\widehat e_i} = {\bf 0}$ from (\ref{eq1}). 
However, for all $i \in {\mathcal{M}_{{\rm{ap}}}} \cup {\mathcal{M}_{{\rm{pa}}}}$, we cannot obtain ${\widehat e_i} = {\bf 0}$ directly from (\ref{eq2}) and (\ref{eq3}). For the manipulator $i \in \mathcal{M}_{\mathrm{ap}}$, note that the virtual end-effector position $\widehat x_i^\text{end}$ relies solely on its active joint angle $q_{i,1}$. Thus, the chain rule in differential calculus gives
\begin{equation}
\frac{{\partial V(e)}}{{\partial {{\widehat x}_{i,X}}^\text{end}}}\frac{{\partial {{\widehat x}_{i,X}}^\text{end}}}{{\partial {q_{i,1}}}} = \frac{{\partial V(e)}}{{\partial {{\widehat x}_{i,Y}}^\text{end}}}\frac{{\partial {{\widehat x}_{i,Y}}^\text{end}}}{{\partial {q_{i,1}}}}, \label{deri}
\end{equation}
which can be rewritten into 
\begin{equation}
{{\bar J}_{i,1}(q_{i,1})}{{\widehat e}_{i,1}} - {{\bar J}_{i,2}(q_{i,1})}{{\widehat e}_{i,2}} = 0. \label{supply}
\end{equation}
Combining (\ref{eq2}) and (\ref{supply}) yields 
\begin{equation}
\begin{aligned}
& \bar{J}_i^*\left(q_{i, 1}\right) \widehat{e}_i={\bf 0}, \\
& \bar{J}_i^*\left(q_{i, 1} \right):=\left[\begin{array}{ll}
\bar{J}_{i, 1}\left(q_{i, 1}\right) & \bar{J}_{i, 2}\left(q_{i, 1}\right) \vspace{0.2cm}\\
\bar{J}_{i, 1}\left(q_{i, 1}\right) & -\bar{J}_{i, 2}\left(q_{i, 1}\right)
\end{array}\right].
\end{aligned} \label{eq:inver}
\end{equation}
The matrix ${\bar J_i ^ *(q_{i,1}})\in \mathbb{R}^{2\times 2}$ is invertible if and only if ${\bar J_{i,1}}(q_{i,1}) {\bar J_{i,2}}(q_{i,1}) \ne 0$, that is \vspace{0.1cm} $\displaystyle q_{i, 1}+\beta_i \neq \frac{k \pi}{2}(k \in \mathbb{Z})$. Thus, under {\bf A2}, Eq. \eqref{eq:inver} yields $ {\widehat e_i} = {\bf 0}$ for all $i \in \mathcal{M}_{\mathrm{ap}}$. Based on a similar analysis, we can deduce ${\widehat e_i} = {\bf 0}$ for all $i\in \mathcal{M}_{\mathrm{pa}}$ under {\bf A3}.
Therefore, 
we have ${\widehat e} = {\bf 0}$. Since $ D^\mathrm{T}(\widehat z) \bar B^\mathrm{T}$ is full rank, we deduce directly from \eqref{defined} that $ {e} = {\bf 0}$, thus completing the proof.
\end{proof}

\section{Simulation results}
\label{sec:sim}
We present simulation results in this section using the distance-based method as an illustrative example. Consider a group of $N=4$ two-link manipulators with flexible joints operating in a gravity-free $X-Y$ plane. The manipulators' mechanical parameters are identical to those in \cite[Chapter 5]{XL14}, where $m_{i,1} = 0.7223 \;\rm{kg}$, $m_{i,2} = 1.2963\; \rm{kg}$, $l_{i,1} = 0.1184 \; \rm{m}$, $l_{i,2} = 0.2357 \; \rm{m}$, $L_{i,1} = 0.3 \; \rm{m}$, $L_{i,2} = 0.5 \; \rm{m}$,
$I_{i,1} = 0.0082 \; \rm{kgm^2}$ and $I_{i,2} = 0.0358 \; \rm{kgm^2}$ for $i = 1,2,3,4$. The stiffness of the torsional springs at the joints is $K_i = \left(\begin{smallmatrix}5 & 0 \\ 0 & 5\end{smallmatrix}\right)$.
Consider the desired shape as a square whose side length is $2 \; \rm{m}$. The incidence matrix $B$ of the corresponding
formation graph $\mathcal{G}$ is
\[
B=\left[\begin{array}{ccccc}
1 & 0 & 0 & -1 & 1 \\
-1 & 1 & 0 & 0 & 0 \\
0 & -1 & 1 & 0 & -1 \\
0 & 0 & -1 & 1 & 0
\end{array}\right].
\]
The manipulators’ base locations are  $x_{1}^\text{base} = [0, 0]^{\rm T}$, $x_{2}^\text{base} = [3, 0]^{\rm T}$, $x_{3}^\text{base} = [3, 2]^{\rm T}$ and $x_{4}^\text{base} = [0, 2]^{\rm T}$, respectively. The relative orientation angle between $\Sigma_W$ and $\Sigma_i$ is $\beta_i=0$ for $i = 1, 2, 3, 4$. The manipulators' initial joint angles are $q_1(0) = [-\pi/2, \pi/3]^{\rm T}$, $q_2(0) = [\pi/3, -\pi/3]^{\rm T}$,
$q_3(0) = [\pi/3, -\pi/3]^{\rm T}$ and $q_4(0) = [-\pi/6, -\pi/3]^{\rm T}$ and initial
joint angular velocities are all zero. 

\subsection{The Existence of Solutions}
\label{sec:solutions}
Remark \ref{re:sw1} discusses the relationship between the cardinality of $\mathcal{S}_W^\text{distance}$ and the number of underactuated manipulators in the network. To verify Remark \ref{re:sw1}, we draw the elements in $\mathcal{S}_W^\text{distance}$ based on the following cases.

1) {\it {\bf Case 1} (one of the manipulators is underactuated)}: Manipulators 1–3 are fully-actuated, and manipulator 4 is the AP manipulator. 

2) {\it {\bf Case 2} (two of the manipulators are underactuated)}: Manipulators 1 and 2 are fully-actuated, manipulator 3 is the AP manipulator, and manipulator 4 is the PA manipulator. 

3) {\it {\bf Case 3} (three of the manipulators are underactuated)}: Manipulator 1 is fully-actuated, manipulators 2 and 3 are the AP manipulators, and manipulator 4 is the PA manipulator. 

4) {\it {\bf Case 4} (all four manipulators are underactuated)}: Manipulators 1--3 are the AP manipulators, and manipulator 4 is the PA manipulator.

\begin{figure}
	\centering
\hspace*{-0.2cm}\includegraphics[scale=0.4]{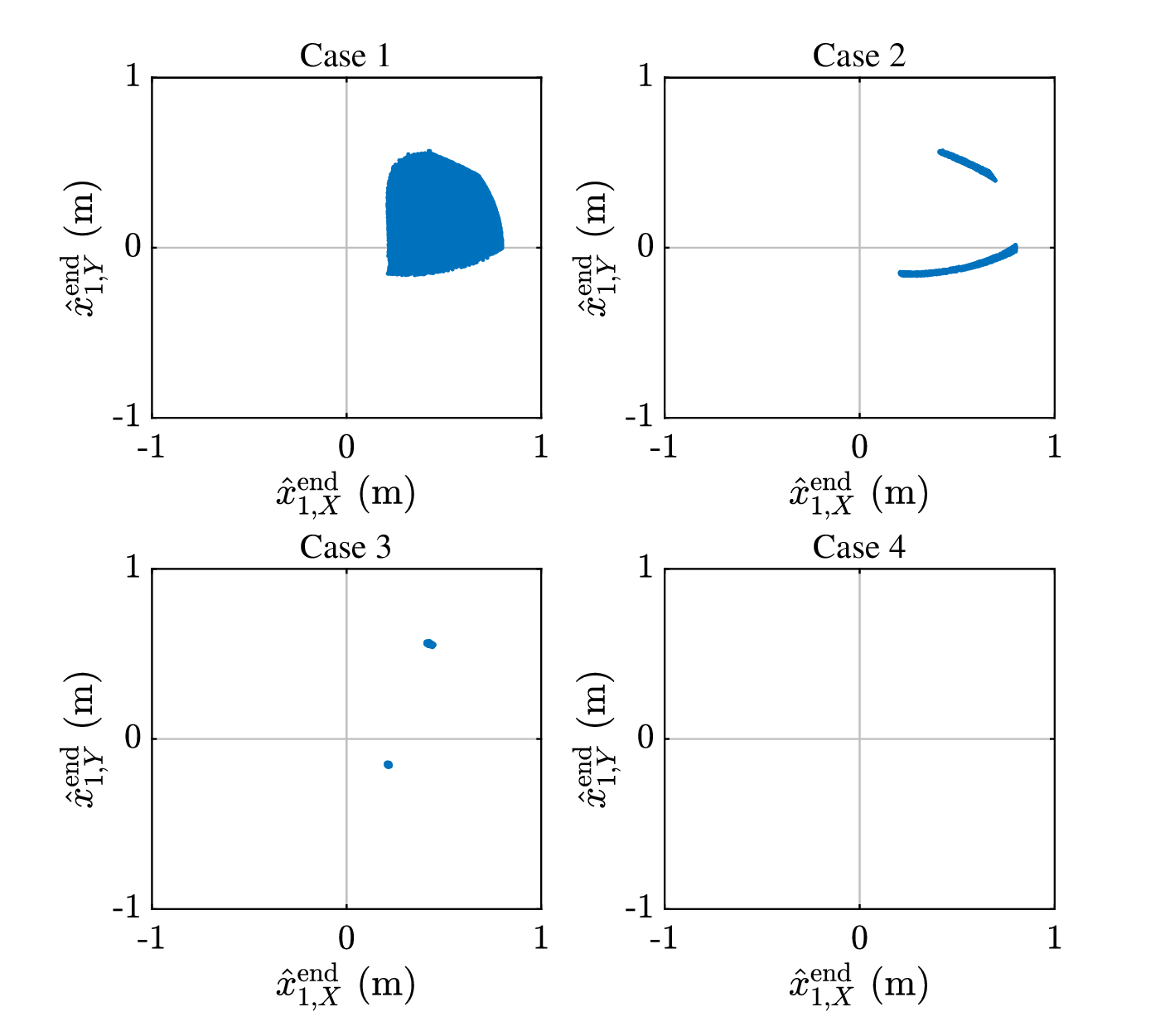}
\caption{The projection of numerically approximated elements in $\mathcal{S}_W^\text{distance}$ onto the $(\widehat x_{1,X}^\text{end}, \widehat x_{1,Y}^\text{end})$ plane for Cases 1--4.}
 \label{solutions}
\end{figure}

We numerically solve the nonlinear system of equations in 
(\ref{z_distance}) by using the \texttt{fsolve} function in MATLAB with the computational accuracies `FunctionTolerance' and `StepTolerance' both set to $1e-8$. The solutions are visualized in the $(\widehat x_{1,X}^\text{end}, \widehat x_{1,Y}^\text{end})$ plane as scatter plots (see Fig.\ref{solutions}).

Fig.\ref{solutions} shows that with an increase in the count of underactuated manipulators, the cardinality of $\mathcal{S}_W^\text{distance}$ decreases. Furthermore, we can make the following conjectures about the projection of the elements in $\mathcal{S}_W^\text{distance}$ onto the $(\widehat x_{1,X}^\text{end}, \widehat x_{1,Y}^\text{end})$ plane:
without computational errors, the elements are distributed in a 2D plane in Case 1; the elements are distributed along curves in Case 2; there are a few isolated elements in Case 3; and there are no elements in Case 4. The obtained results are consistent with the statement in Remark \ref{re:sw1}.

\subsection{Performance of the Controller}
\begin{figure}
	\centering
	\includegraphics[scale=0.4]{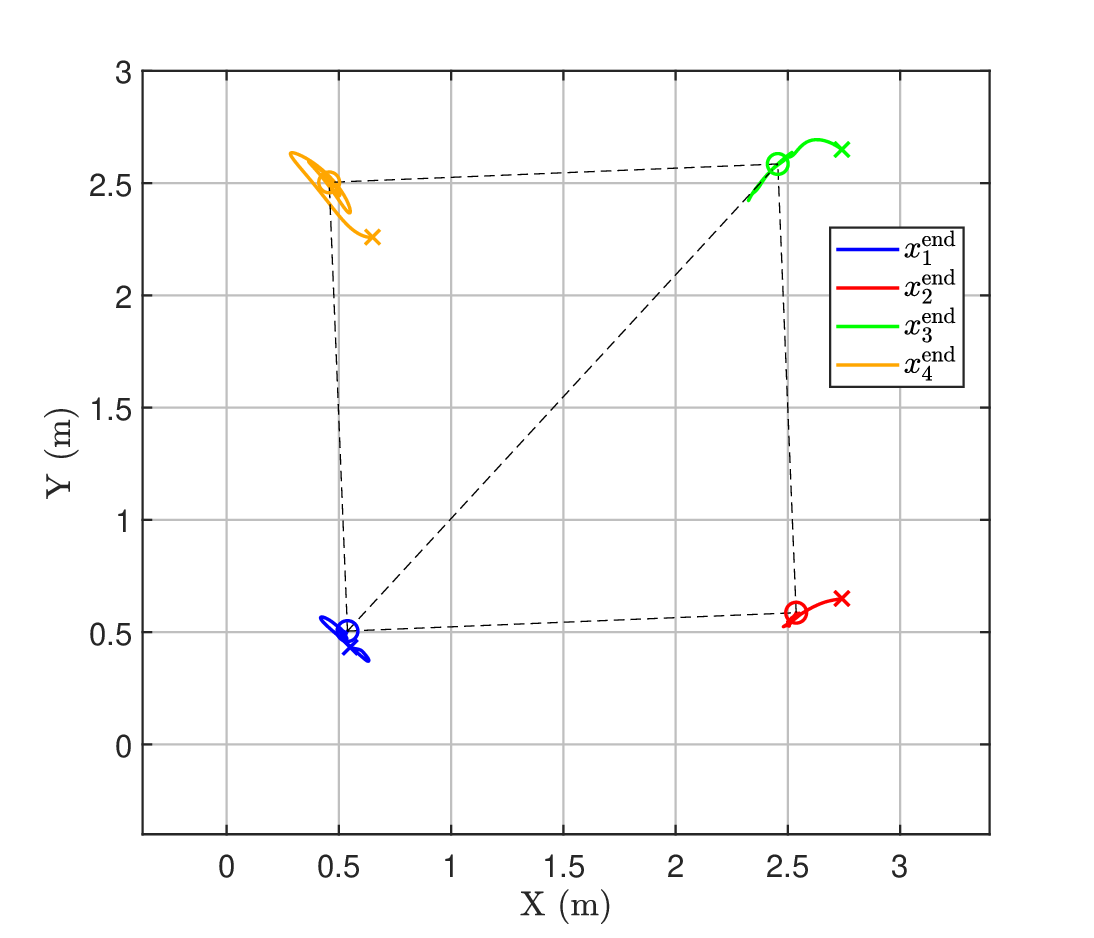}
	\caption{Case 2: trajectories of manipulators' end-effectors from initial ($\times$) to final positions ($\circ$).}
 \label{formation}
\end{figure}

\begin{figure}
	\centering
\includegraphics[scale=0.4]{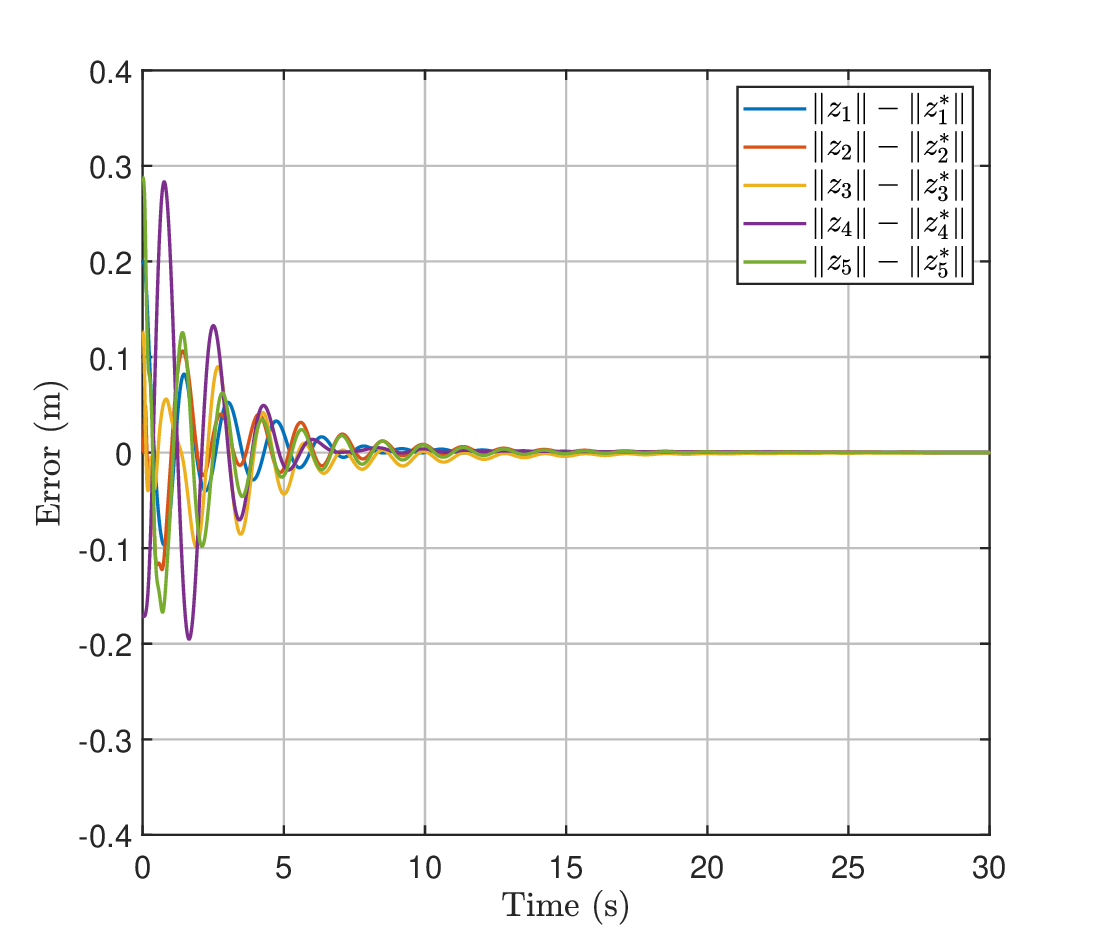}
	\caption{Case 2: performance of the inner distance error between manipulators' end-effectors.}
  \label{distance}
\end{figure}

\begin{figure}
	\centering
\includegraphics[scale=0.4]{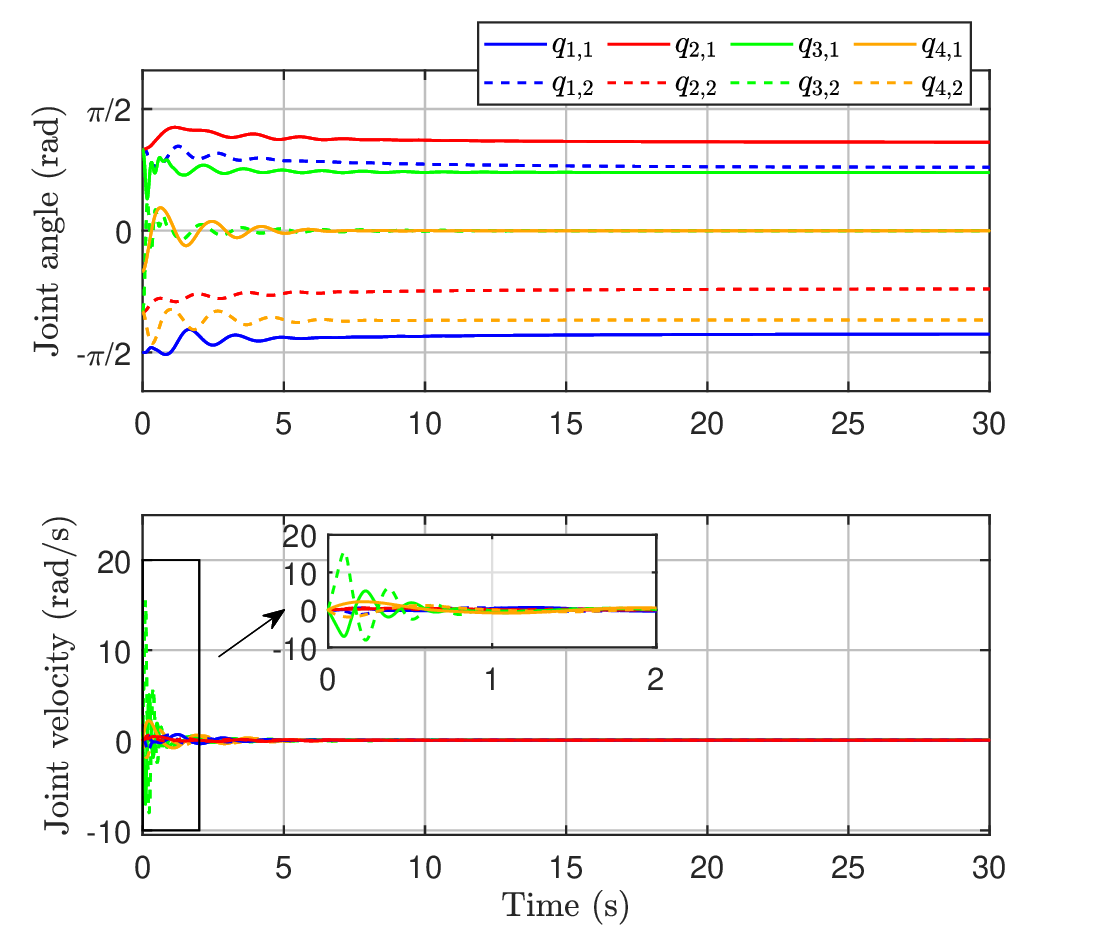}
	\caption{Case 2: manipulators' joint angle and angular velocity signals.}
 \label{fig:states}
\end{figure}

\begin{figure}
	\centering
	\includegraphics[scale=0.4]{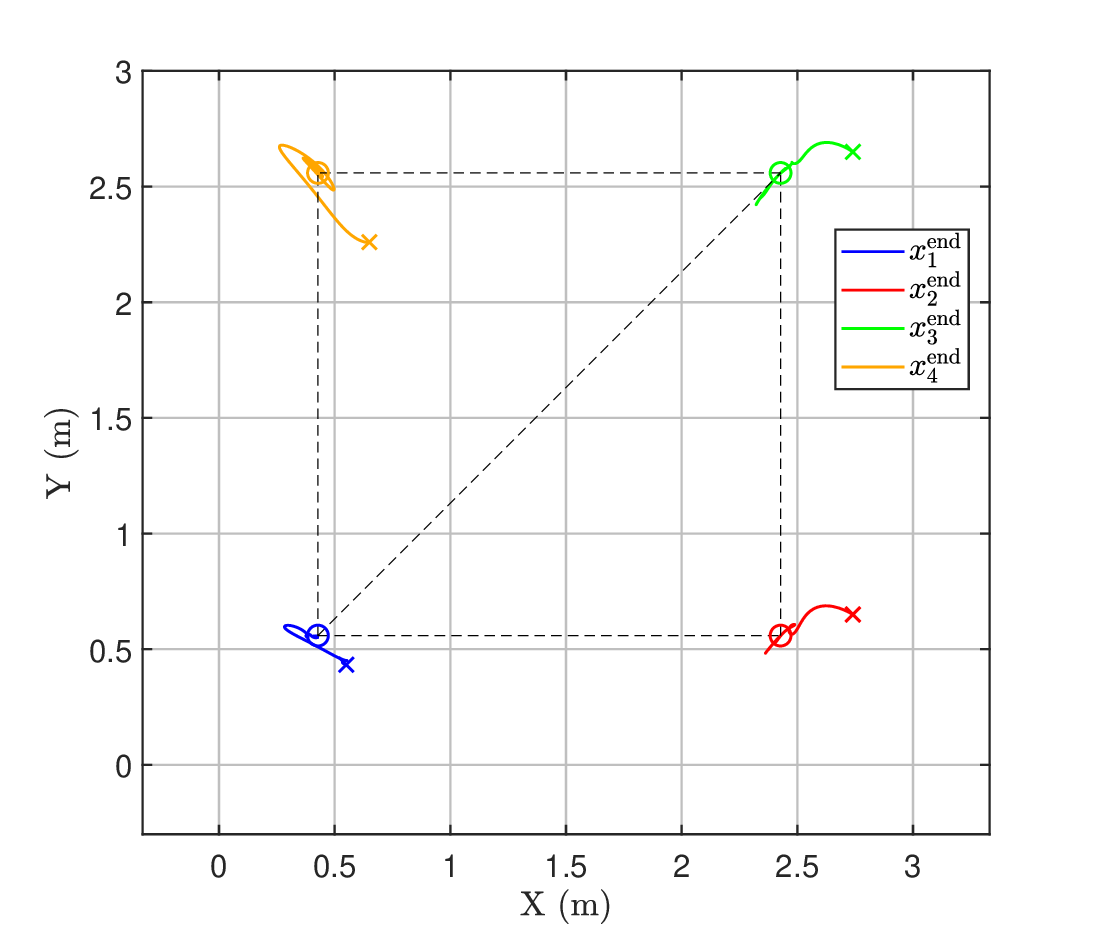}
\caption{Case 3: trajectories of manipulators' end-effectors from initial ($\times$) to final positions ($\circ$).}
 \label{formation_4_3}
\end{figure}

\begin{figure}
	\centering
	\includegraphics[scale=0.4]{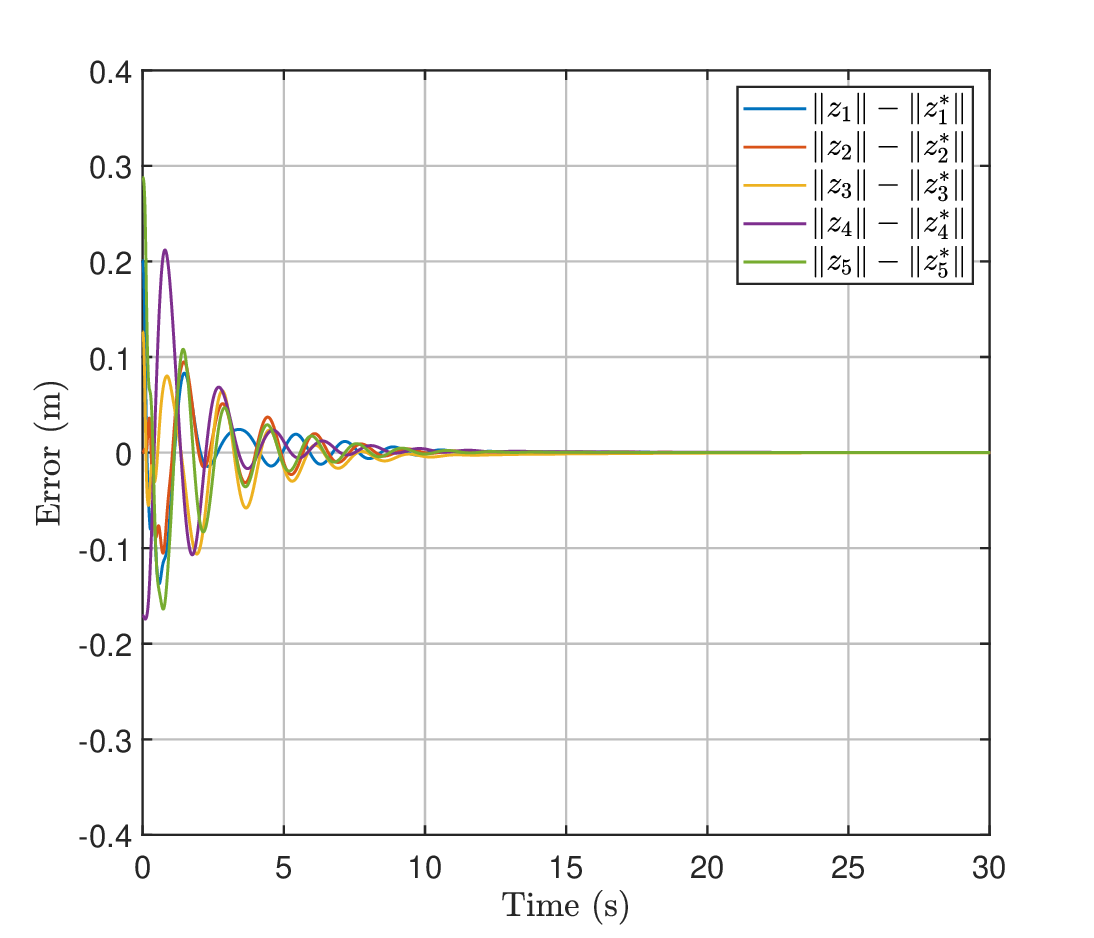}
\caption{Case 3: performance of the inner distance error between manipulators' end-effectors.}
  \label{distance_4_3}
\end{figure}

We take Cases 2 and 3 as examples to verify the effectiveness of the proposed controller (\ref{eq:controller}). For Case 2, we set $k_S = 0.5I_5$, $k_D = 0.4$, and the result is shown in Figs. \ref{formation}--\ref{fig:states}. Fig. \ref{formation} depicts the trajectories of the manipulators' end-effectors, while Fig. \ref{distance} illustrates the convergence of inner distances between end-effectors to desired values. Figs. \ref{formation} and \ref{distance} show that the end-effectors eventually achieve the desired shape. From Fig. \ref{fig:states}, we notice that the manipulators' joint angles converge to constants.  
Moreover, we notice that for underactuated manipulators, the passive joint angles $q_{3,2}, q_{4,1}$ converge to 0, which means that $x_i^\text{end}$ converges to $\widehat x_i^\text{end}$. For Case 3, we set $k_S = 0.4I_5$, $k_D = 0.5$, and the results in Figs. \ref{formation_4_3} and \ref{distance_4_3} show that the proposed controller (\ref{eq:controller}) is also
effective. Furthermore, from Figs. \ref{formation} and \ref{formation_4_3}, we notice that the final position of $x_1^\text{end}$ belongs to the point set depicted in Fig. \ref{solutions}.

\section{Conclusion}
\label{sec:con}
This paper investigates the issue of end-effector formation keeping for networked two-link manipulators with flexible joints operating in the same gravity-free plane. We extend two existing distributed formation control strategies, namely distance-based and displacement-based methods, to the case where some of these manipulators are underactuated. The distance-based method is shown to be effective for the group with three or fewer underactuated manipulators, while the displacement-based method is effective for two or fewer. 

\bibliographystyle{plain}        
\bibliography{pzy}           

\appendix

\section{Proof of Lemma~\ref{lemma_AP}}
\label{app A}
For the AP manipulator $i \in \mathcal{M}_{\rm ap}$ with flexible joints, putting $q_{i,1} \equiv q_{i,1}^*$ and $u_{i,1} \equiv u_{i,1}^*$ (where $q_{i,1}^*,u_{i,1}^* $ are constants) into (\ref{full}) yields
\begin{align}
&\begin{aligned}
& \left(\alpha_{i, 2}+\alpha_{i, 3} \cos q_{i, 2}\right) \ddot{q}_{i, 2}-\alpha_{i, 3} \dot{q}_{i, 2}^2 \sin q_{i, 2} \\
& =u_{i, 1}^*-K_{i, 1} q_{i, 1}^*=\lambda_1, \label{AP_1}
\end{aligned} \\
&\begin{aligned}
{\alpha _{i,2}}{{\ddot q}_{i,2}} + {K_{i,2}}{q_{i,2}} = 0, \label{AP_2}
\end{aligned}
\end{align}
where $\lambda_1$ is a constant. Since $\dot q_{i,2}$ is bounded, multiplying both sides of (\ref{AP_2}) by $\dot q_{i,2}$ yields
\begin{equation}
{\alpha _{i,2}}{{\ddot q}_{i,2}}{{\dot q}_{i,2}} + {K_{i,2}}{q_{i,2}} {{\dot q}_{i,2}}=0. \label{AP_3}
\end{equation}
Respectively, integrating \eqref{AP_1} and (\ref{AP_3}) with respect to time $t$ yields 
\begin{align}
&\begin{aligned}
\left( {{\alpha _{i,2}} + {\alpha _{i,3}}\cos {q_{i,2}}} \right){{\dot q}_{i,2}} = {\lambda _1}t + {\lambda _2}, \label{AP_1_inte1}
\end{aligned}\\
&\begin{aligned}
{\alpha _{i,2}}\dot q_{i,2}^2 + {K_{i,2}}q_{i,2}^2 = {\lambda_3}, \label{AP_3_inte}
\end{aligned}
\end{align}
where $\lambda_2, \lambda_3$ are constants. Given the boundness of $\dot q_{i,2}(t)$, the right-hand side of (\ref{AP_1_inte1}) is bounded for all $t$. Therefore, we have $\lambda_1=0$, that is, the sum torque (due to the torsional spring and the active control torque) at the first joint is zero. Then, by integrating (\ref{AP_1_inte1}) further with respect to time $t$, we obtain
\begin{equation}
{\alpha _{i,2}}{q_{i,2}} + {\alpha _{i,3}}\sin {q_{i,2}} = {\lambda _2}t + {\lambda _4}, \label{AP_1_inte2_inte}
\end{equation}
where $\lambda_4$ is a constant. Since $\dot q_{i,2}$ is bounded, we know that $q_{i,2}(t)$ is also bounded from (\ref{AP_3_inte}). Consequently, we know that the right-hand side of (\ref{AP_1_inte2_inte}) remains bounded for all $t$, leading us to conclude $\lambda_2=0$. Subsequently, from (\ref{AP_1_inte1}), we have $\ddot q_{i,2}=\dot q_{i,2}=0$. This allows us to deduce $q_{i,2}=0$ from (\ref{AP_2}), thus completing the proof.

\section{Proof of Lemma~\ref{lemma_PA}}
\label{app B}
For the PA manipulator $i \in \mathcal{M}_{\rm pa}$ with flexible joints, putting $q_{i,2} \equiv q_{i,2}^*$ and $u_{i,2} \equiv u_{i,2}^*$ (where $q_{i,2}^*,u_{i,2}^* $ are constants) into (\ref{full}) yields
\begin{align}
&\begin{aligned}
& M_{i, 11}\left(q_{i, 2}^*\right) \ddot{q}_{i, 1}+K_{i, 1} q_{i, 1}=0,  \label{PA_1}
\end{aligned} \\
&\begin{aligned}
M_{i, 21}\left(q_{i, 2}^*\right) \ddot{q}_{i, 1}+\alpha_{i, 3} \dot{q}_{i, 1}^2 \sin q_{i, 2}^*=u_{i, 2}^*-K_{i, 2} q_{i, 2}^*.  \label{PA_2}
\end{aligned}
\end{align}
Since $M_i(q_i)$ is positive definite, we have $M_{i,11}\left(q_{i,2}^*\right)>0$. Therefore, we can eliminate $\ddot q_{i,1}$ in (\ref{PA_2}) by using (\ref{PA_1}), and
\begin{align}
\frac{-K_{i, 1} M_{i, 21}\left(q_{i, 2}^*\right) q_{i, 1}}{M_{i, 11}\left(q_{i, 2}^*\right)}+\alpha_{i, 3} \dot{q}_{i, 1}^2 \sin q_{i, 2}^*=u_{i, 2}^*-K_{i, 2} q_{i, 2}^*. \label{PA_3}
\end{align}
Taking the time derivative of (\ref{PA_3}) yields
\begin{align}
\frac{{ - {K_{i,1}}{M_{i,21}}\left( {q_{i,2}^*} \right){{\dot q}_{i,1}}}}{{{M_{i,11}}\left( {q_{i,2}^*} \right)}} + 2{\alpha _{i,3}}{{\dot q}_{i,1}}{{\ddot q}_{i,1}}\sin q_{i,2}^* = 0. \label{PA_4}
\end{align}
Eliminating $\ddot q_{i,1}$ in (\ref{PA_4}) with (\ref{PA_1}) yields
\begin{align}
{{\dot q}_{i,1}}{K_{i,1}}\left\{ {{M_{i,21}}\left( {q_{i,2}^*} \right) + 2{\alpha _{i,3}}{q_{i,1}}\sin q_{i,2}^*} \right\} = 0. \label{PA_5}
\end{align}
Now, we show that if $\alpha_{i,2} \ne \alpha_{i,3}$, then $\dot q_{i,1}=0$. Otherwise, if $\dot q_{i,1} \ne 0$, (\ref{PA_5}) is reduced to 
\begin{align}
{\alpha _{i,2}} + {\alpha _{i,3}}\cos q_{i,2}^* + 2{\alpha _{i,3}}{q_{i,1}}\sin q_{i,2}^* = 0. \label{PA_6}
\end{align}
Since $\dot q_{i,1} \ne 0$, the left-hand side of (\ref{PA_6}) is constant only if $\sin q_{i,2}^*=0$. So, we can reduce (\ref{PA_6}) to $\alpha_{i,2} \pm \alpha_{i,3}=0$, which introduces a contradiction with $\alpha_{i,2} \ne \alpha_{i,3}$ and the fact that $\alpha_{i,2}$ and $\alpha_{i,3}$ are both positive. Since $\ddot q_{i,1}=\dot q_{i,1}=0$, we have $q_{i,1}=0$ from (\ref{PA_1}).
\end{document}